\documentclass[journal,10pt]{IEEEtran}

\usepackage{amsmath,amssymb,amsfonts,graphicx,mathtools,nccmath,amsthm,float,stfloats}
\usepackage{multicol,tikz,cite,array,booktabs,url}
\usepackage{cite}
\usepackage{textcomp}
\usepackage{xcolor, flushend}
\usepackage{algorithm}
\usepackage{algpseudocode,algorithmicx}
\usepackage[top=.75in, left= 0.625in, right = 0.63in, bottom =1in]{geometry}

\algnewcommand{\LineComment}[1]{\State \(\#\) #1}
    
\graphicspath{{figs/}}
\DeclareGraphicsExtensions{.pdf}

\newtheorem*{remark}{Remark}

\DeclareMathOperator{\diag}{diag}
\DeclareMathOperator{\atan}{atan}

\DeclareMathOperator{\trace}{Tr}
\newcommand{\herm}{\mathrm{H}}
\newcommand{\tran}{\mathrm{T}}
\renewcommand{\Re}{\mathrm{Re}}
\renewcommand{\Im}{\mathrm{Im}}
\newcommand\norm[1]{\lVert#1\rVert}

\algnewcommand\algorithmicinput{\textbf{Set}}
\algnewcommand\Set{\item[\algorithmicinput]}
\algnewcommand\algorithmicinitial{\textbf{Initialize}}
\algnewcommand\Initialize{\item[\algorithmicinitial]}

\let\oldReturn\Return
\renewcommand{\Return}{\State\oldReturn}

\begin{document}

\title{Active Reconfigurable Intelligent Surfaces: Circuit Modeling and Reflection Amplification Optimization}
\author{Panagiotis~Gavriilidis,~\IEEEmembership{Graduate Student Member,~IEEE}, Deepak Mishra,~\IEEEmembership{Senior Member,~IEEE}\\ Besma Smida,~\IEEEmembership{Senior Member,~IEEE}, Ertugrul Basar,~\IEEEmembership{Fellow,~IEEE}, Chau~Yuen,~\IEEEmembership{Fellow,~IEEE},\\ and George C. Alexandropoulos,~\IEEEmembership{Senior Member,~IEEE} 
\thanks{This work has been supported by the SNS JU project TERRAMETA under the European Union’s Horizon Europe research and innovation programme under grant agreement number 101097101, including top-up funding by UKRI under the UK government's Horizon Europe funding guarantee.}
\thanks{P. Gavriilidis and G. C. Alexandropoulos are with the Department of Informatics and Telecommunications,
National and Kapodistrian University of Athens, 16122 Athens, Greece. G. C. Alexandropoulos is also with the Department of Electrical and Computer Engineering, University of Illinois Chicago, IL 60601, USA (e-mails: \{pangavr, alexandg\}@di.uoa.gr).}
\thanks{D. Mishra is with the School of Electrical Engineering and Telecommunications, University of New South Wales, NSW 2052 Sydney, Australia (e-mail: d.mishra@unsw.edu.au).}
\thanks{B. Smida is with the Department of Electrical and Computer Engineering, University of Illinois Chicago, IL 60601, USA (e-mail: smida@uic.edu).}
\thanks{E. Basar is with the Department of Electrical Engineering, Tampere University, 33720 Tampere, Finland, on leave from the Department of Electrical and Electronics Engineering, Ko\c{c} University, 34450 Sariyer, Istanbul, Turkey (e-mails: ertugrul.basar@tuni.fi, ebasar@ku.edu.tr).}
\thanks{C. Yuen is with the School of Electrical and Electronics Engineering, Nanyang Technological University, Singapore (e-mail: chau.yuen@ntu.edu.sg).}
}

\maketitle

\begin{abstract}
Reconfigurable Intelligent Surfaces (RISs) constitute a promising emerging technology that enables wireless systems to control the propagation environment to enhance diverse communication objectives. To mitigate double-fading attenuation in RIS-aided links, the paradigm of active metamaterials capable of amplifying their incident wave has emerged. In this paper, capitalizing on the inherent negative-resistance region of tunnel diodes, we propose their integration into each RIS unit element to enable RISs with reflection amplification entirely in the analog domain. We derive novel realistic phase-amplitude relationships and power constraints specific to this model, addressing gaps in the existing literature where amplitude limits are often chosen arbitrarily. This characterization of our active RIS unit elements is incorporated into two novel optimization frameworks targeting the spectral efficiency maximization of RIS-assisted Multiple-Input-Multiple-Output (MIMO) systems, which are solved via an one-step approach and an iterative Alternating Optimization (AO) method. The former approach is used to initialize the AO framework, enhancing both its performance and convergence. Our numerical investigations emphasize the importance of accurately modeling phase-amplitude dependencies, and provide key insights into the impact of RIS-induced noise as well as the trade-off between available power and the number of active elements.
\end{abstract}


\section{Introduction} \label{Sec:Intro}
Reconfigurable Intelligent Surfaces (RISs)~\cite{George_Basar_RIS_magazine} are envisioned as a revolutionary means to transform any passive wireless communication environment to an active reconfigurable one~\cite{EURASIP_RIS}, offering increased environmental intelligence for diverse communication objectives~\cite{9864655}. An RIS is an artificial planar structure with integrated electronic circuits \cite{WavePropTCCN} that can be programmed to manipulate incoming ElectroMagnetic (EM) fields in a wide variety of functionalities \cite{Marco_Visionary_2019}. Specifically, the RIS unit elements can alter not only the phase of EM waves, but also the magnitude, frequency, and even polarization~\cite{HMIMO}. This level of control enables a wide range of applications, including index~\cite{10247149} and reflection~\cite{9217944} modulation, polarization shift keying~\cite{ibrahim2022binary}, frequency-selective beamforming~\cite{8064681}, energy harvesting~\cite{10693440}, and integrated sensing and communications~\cite{10243495}.

The most prominent function of RISs is the reconfiguration of the reflection responses of their unit elements; such RISs are classified as passive RISs~\cite{George_Basar_RIS_magazine}. This functionality has been extensively studied in the literature~\cite{Jian_RIS_survey}, and the primary building blocks of passive RIS prototypes are their electronically controlled metamaterials, such as semiconductor devices (e.g., PIN and varactor diodes) and MEMS \cite{RIS_prototypes_and_power_consumption_model,abadal2020programmable,EURASIP_RIS}.
These mechanisms are particularly attractive due to their high switching speeds, on the order of microseconds or nanoseconds \cite{amri2022recent}. However, the performance of passive RISs is fundamentally limited by multiplicative fading effects. To address this issue, active RISs, which enable controllable amplification of incident waves, have gained increasing research interest \cite{ Act_vs_Pass, MIMO_active_RIS_robust_BF, active_RIS_survey}. In particular, negative resistance components, such as Tunnel Diodes (TDs), can be used to amplify signals directly in the Radio Frequency (RF) domain, eliminating the need for down-conversion, thus, maintaining a relatively low power budget~\cite{Active_Scatterer_2012, Tunneling_RFID_2018}. These active load elements have also been successfully employed in the past in backscatter communication systems~\cite{Tag_Backscatter_2014, Active_Backscatter_2019}.


To our best of knowledge, the performance of an active RIS was first studied in \cite{Larsson_2021}, where an Alternating Optimization (AO) algorithm was presented to maximize the Signal-to-Noise Ratio (SNR) in an RIS-aided single-input multiple-output communication system. Theoretical limits were derived to characterize the trade-off between the number of active RIS elements and the available power for amplification. Later, \cite{Act_vs_Pass} explored the asymptotic limits of the SNR gain for passive and active RISs, identifying the threshold in terms of the number of RIS elements, \(N\), beyond which a passive RIS outperforms an active one. The SNR gain was shown to scale quadratically with \(N\) for a passive RIS (i.e., \(N^2\)), whereas it scales linearly for an active RIS (i.e., \(N\)). However, it was demonstrated that a passive RIS outperforms an equally sized active one only at extreme scales, which translates to \(N\) being on the order of millions in typical scenarios. Furthermore, the authors in~\cite{Power_Budget} provided a theoretical comparison between active and passive RISs under identical energy constraints, concluding that a passive RIS is superior only in extreme scenarios, such as when the power budget is minimal or $N$ is prohibitively large. 

Active RISs have exhibited significant potential in various use cases beyond SNR maximization, particularly, in wireless-powered~\cite{8369144} and secure~\cite{9501003} communication systems. For the former use case,~\cite{SWIPT} proposed an active RIS-aided simultaneous wireless information and power transfer scheme, showcasing the capability of RISs to effectively extend the range of wireless power transfer. Similarly, the authors in~\cite{zeng2022throughput} investigated a wireless-powered communication use case, revealing that active RISs enhance energy harvesting at wireless-powered devices compared to their passive counterparts, while also providing insights into optimal RIS deployment locations. In secure communication systems, active RISs have proven advantageous across multiple performance metrics. Recent studies have highlighted their superiority in improving the secrecy rate~\cite{SecrecyBai}, reducing the probability of secrecy outage~\cite{Khoshafa_2021}, and even improving energy efficiency. Notably, active RISs can achieve performance superior to passive RISs while requiring fewer elements, thereby, addressing scalability challenges~\cite{Secrecy_Green}. Very recently, in~\cite{10636047}, active RISs were optimized to realize amplitude phase shift keying modulation in RIS-aided cooperative backscatter communications.


Fully active RISs, where all their elements possess amplifying capabilities, is not always the optimal choice due to potential power consumption constraints. To address this, hybrid passive-active RIS designs, where only a subset of elements has amplifying capabilities, have been explored in the literature~\cite{optimization_of_number_of_active_elements,HybrAct}. For instance, a multiuser communication system assisted by a hybrid passive-active RIS was investigated in~\cite{HybrAct}. By jointly optimizing transmit beamformers and the responses of the hybrid RIS unit elements, significant performance improvements over a fully passive RIS, even with only \(4\) out of \(50\) elements being active, was demonstrated. Additionally, \cite{SubConActive} introduced a sub-connected architecture, where groups of elements share a common active load, achieving increased energy efficiency compared to fully active RISs. In \cite{altunbasAct}, this sub-connected architecture was applied to a novel paradigm called hybrid reflection modulation, where RIS-side amplification enables amplitude modulation on the incident wave without requiring baseband signal processing. Lastly, a distinct active RIS framework comprising two passive RISs connected by an amplifier was proposed in~\cite{George_Active}, highlighting notable improvements in throughput and error performance compared to passive RISs.

\subsection{Motivation and Contributions}
Although, a TD for implementing active metamaterials was proposed for active RISs in~\cite{Larsson_2021}, phase-amplitude-dependent modeling was not considered. In addition, the power constraint applied was that of a typical power amplifier, which is unsuitable for TDs. Furthermore, the existing literature often lacks physical constraints on the achievable range of amplification for active metamaterial elements, with maximum amplitude values being arbitrarily chosen \cite{active_RIS_survey}. In this paper, we address these gaps by studying the TD model from a circuit perspective, deriving realistic amplification limits, and retrieving the power consumption model directly from the diode's current-voltage (I-V) characteristics. We also build on the transmission line circuit model of~\cite{Abeywickrama_2020}, which captures practical phase-dependent amplitude variations in passive RIS unit elements, and extend it to active elements. Our active RIS model is then  considered for RIS-assisted MIMO communications where we propose an optimization framework for jointly designing the transmitter's (TX) precoding matrix, the receiver's (RX) combining matrix, and the configuration of the active RIS to maximize the achievable spectral efficiency. The contributions of this paper are summarized as follows:  
\begin{itemize}  
    \item We introduce a new model for the reflection amplification coefficient of TD-based active RIS unit elements that combines the transmission line circuit model with the TD's characteristics. 
    We study the phase-amplitude dependency in this model, deriving the amplitude limits of the reflection coefficient as a function of the imposed phase shift, and determine the circuit parameters that correspond to the minimum and maximum amplitude cases.
    \item Considering circuit constraints, we derive the feasible range of negative resistance values for our TD-based active RIS unit element implementation, which in turn helps us to identify the feasible amplitude range of the reflection amplification coefficient, thus, overcoming the arbitrary amplitude constraints commonly found in the active RIS open technical literature.  
    \item Using the I-V TD characteristics, we present a power consumption model that offers a more accurate representation compared to conventional amplifier-based models in the state-of-the-art active RIS works.
    \item A novel approximate expression for the RIS reflection vector is introduced that incorporates both active and passive unit elements as well as phase-amplitude dependency.
    \item Two optimization approaches for the maximization of the achievable spectral efficiency in active-RIS-aided MIMO communication systems are presented. The first approach adheres to the derived constraints on the active RIS and is based on AO, whereas the second one is an one-step scheme that serves either as a standalone approach or as an initialization step to enhance the performance and convergence of the AO-based approach.   
    \item Our extensive numerical investigations demonstrate the superiority of our optimized active RIS designs for MIMO communications over benchmarks, providing key insights into the impact of active-RIS-generated noise as well as on the trade-off between the number of active RIS elements and the available power budget.  

\end{itemize}  

The remainder of the paper is organized as follows. Section~\ref{sec:RIS with Active Elements} introduces our phase-amplitude dependence model, describes the correspondence between the reflection amplification coefficient and circuit parameters, and formulates our power consumption model. Section~\ref{sec:System Model and Problem Formulation} presents the end-to-end system model for active-RIS-aided MIMO communications, as well as the optimization framework with design parameters the TX precoding matrix, the RX combining matrix, and the configuration of the active RIS. Section~\ref{sec:solution_methodologies} outlines the proposed solution methodologies, whose performance evaluation results are are provided in Section~\ref{sec: Simulation Results}, demonstrating the advantages of the proposed approach compared to benchmarks. Finally, the concluding remarks of the paper are drawn in Section~\ref{sec: Conclusion}.  

\textit{Notations:} Vectors and matrices are denoted by boldface lowercase and boldface capital letters, respectively. The transpose, conjugate, Hermitian transpose, trace, and inverse of $\mathbf{A}$ are denoted by $\mathbf{A}^T$, $\mathbf{A}^*$, $\mathbf{A}^H$, ${\rm Tr}\{\mathbf{A}\}$, and $\mathbf{A}^{-1}$ respectively, while $\mathbf{I}_{n}$ and $\mathbf{0}_{n\times 1}$ ($n\geq2$) are the $n\times n$ identity matrix and $n\times 1$ zeros' vector, respectively. $[\mathbf{A}]_{i,j}$ is the $(i,j)$-th element of $\mathbf{A}$, $[\mathbf{a}]_i$ is $\mathbf{a}$'s $i$-th element, ${\rm diag}\{\mathbf{a}\}$ denotes a square diagonal matrix with $\mathbf{a}$'s elements in its main diagonal. The operand \(\odot\) represents the Hadamard matrix product, while the operand \(\otimes\) is the Kronecker matrix product. Furthermore, \({\rm vec}(\mathbf{A})\) denotes the vectorized form of matrix \(\mathbf{A}\) by stacking its columns in order. Moreover, \([a]^+\) denotes the operation \(\max(0,a)\). $\mathbb{R}$ and $\mathbb{C}$ represent the sets of real and complex numbers, respectively, $|a|$ and ${\rm arg}(a)$ are respectively the amplitude and phase of the complex number $a$, while $\Re[a]$ and $\Im[a]$ return $a$'s real and imaginary parts, respectively, and \(\norm{\cdot}_{\rm F}\) denotes the Frobenius norm of the involved matrix/vector. Additionally, \(\nabla_{\mathbf{x}}f(\mathbf{x})\) denotes the Euclidean gradient of the scalar function \(f(\cdot)\) with respect to \(\mathbf{x}\). The expression $\mathbf{x}\sim\mathcal{CN}(\mathbf{a},\mathbf{A})$ indicates a complex Gaussian random vector with mean $\mathbf{a}$ and covariance matrix $\mathbf{A}$, and $\jmath\triangleq\sqrt{-1}$ is the imaginary unit. Finally, \(\atan(\cdot)\) represents the inverse tangent function.

\section{RIS Reflection Amplification Modeling}\label{sec:RIS with Active Elements}
The typical hardware implementation of RISs is based on two-dimensional metamaterials, which are digitally controllable by printed circuit boards \cite{cui2014coding}. RIS unit elements/cells are widely modeled as resonant circuits, enabling modification of the reflection amplitude and phase of incident EM waves through proper design of their circuit variables. To model these elements, we adopt the transmission line circuit model, according to which each $n$-th RIS unit element (with $n=1,\ldots,N$) behaves like a parallel resonant circuit, and its impedance is given by \cite[eq.~(3)]{Abeywickrama_2020}:
\begin{equation}\label{eq:Z_n}
   Z_n(C_n, R_n)\triangleq\frac{\jmath\omega L_1(\jmath\omega L_2+\frac{1}{\jmath\omega C_n} +R_n)}{\jmath\omega L_1+(j\omega L_2+\frac{1}{\jmath\omega C_n} +R_n)},  
\end{equation}  
where $L_1$ and $L_2$ denote the inductances of the bottom and top layers, respectively, and $\omega$ is the angular frequency of the incident signal, while $C_n$ and $R_n$ represent the tunable capacitance and tunable resistance, respectively. Let \(\boldsymbol{\gamma} \in \mathbb{C}^N\) denote the reflection coefficient vector of an \(N\)-element RIS, where each element's reflection coefficient \([\boldsymbol{\gamma}]_n\) results from the impedance discontinuity between free space (\(Z_0 = 377\,\Omega\)) and the element's impedance \(Z_n(C_n, R_n)\), and is given by:
\begin{equation}\label{eq:ref}
   [\boldsymbol{\gamma}]_n(C_n,R_n) \triangleq \frac{Z_n(C_n, R_n) -Z_0}{Z_n(C_n, R_n) +Z_0}.
\end{equation}
Clearly, the value of this coefficient (i.e., both its amplitude $\alpha_n$ and phase $\varphi_n$) is a function of the tunable $R_n$ and $C_n$, whose adjustment may yield various phase/reflection configurations.

\subsection{Reflection Amplification via Negative Resistance}
\begin{figure}[t]
 \includegraphics[width=\columnwidth]{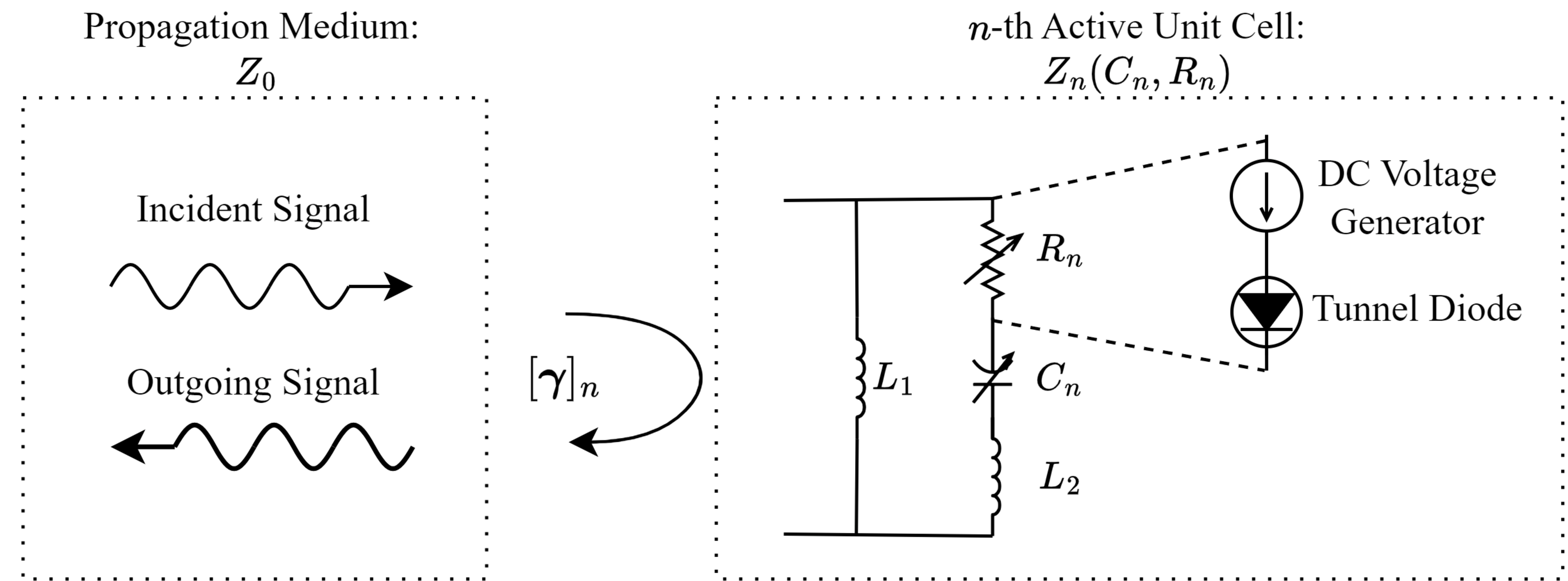}
 \caption{The tunnel diode embedded in~\cite{Abeywickrama_2020}'s transmission line circuit model for each \(n\)-th RIS unit element, with \([\boldsymbol{\gamma}]_n\) denoting its reflection coefficient.}
 \label{fig:RIS_atom}
 \end{figure}
Many RIS implementations restrict the resistance \(R_n\) to non-negative values, assuming it accounts for power dissipation due to losses in semiconductor devices, metals, and dielectrics, which cannot be eliminated in practice~\cite{PhysRevApplied.11.044024,Abeywickrama_2020}. However, the impedance of each \(n\)-th element in~\eqref{eq:ref} can be extended to an active load considering a negative value for \(R_n\), which consequently enables tunable reflection amplification. 
Negative resistance is a property where the current through a device flows opposite to the applied voltage, or the voltage generated opposes the given current \cite{amp1}. TDs have been widely used to achieve negative resistance, mainly due to their low cost and minimal Direct Current (DC) power consumption. Capitalizing on the TD design in~\cite{8057976}, we propose embedding a TD in each RIS unit element to enable tunable reflection amplification, as illustrated in Fig.~\ref{fig:RIS_atom}. Essentially, the now negative resistance \(R_n\) is implemented with a TD driven by a DC voltage generator. 

It is noted that, in practice, an amplifier's negative resistance operation is constrained to a specific range of input power, referred to as the dynamic region, as illustrated in \cite{8057976} and \cite[Figs. 3 and 5]{nonlinear}.
In this paper, we assume that the power of the incoming signal in our TD-based active RIS unit element is small enough to keep the amplification within its linear domain, since amplification outside this range is unnecessary~\cite{Larsson_2021}.

\subsection{Tunnel Diode (TD) Modeling}\label{tunnel_diode_model}
\begin{figure}[t]
    \centering
    \includegraphics[width=0.8\columnwidth]{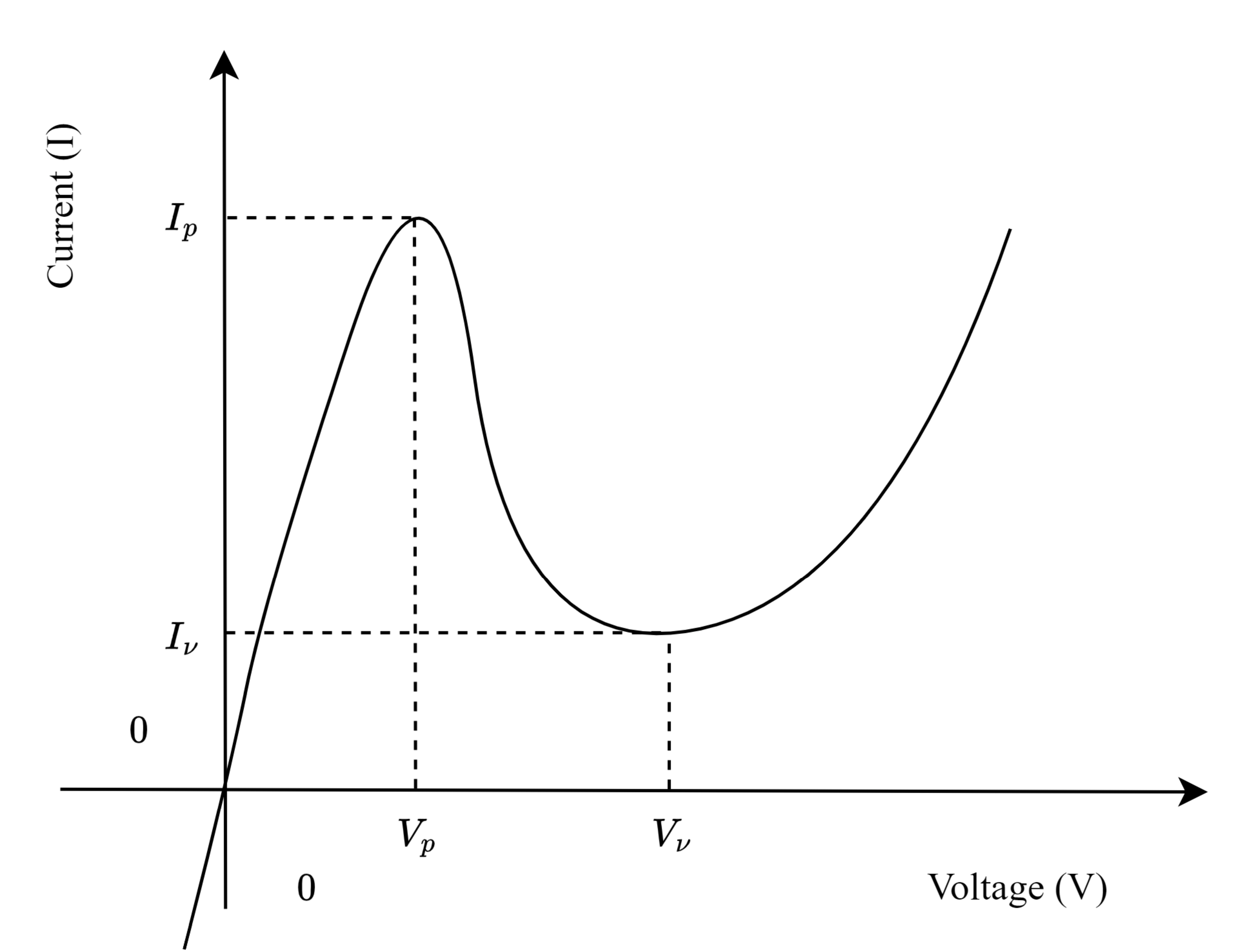}
    \caption{The I-V characteristic of a typical tunnel diode. The operating region for the considered TD-based negative resistance lies between \(V_p\) and \(V_{\nu}\), which represent the peak and valley voltages, respectively.}
    \label{fig:I-V characteristic}
\end{figure}

The TD current is typically characterized via the following three terms: \textit{i}) the current in the tunneling region \(I_T(V)\) for voltages $V$ applied before the valley voltage \(V_{\nu}\), i.e., for \(V<V_{\nu}\); \textit{ii}) the current after the tunneling valley \(I_d(V)\) (diffusion current), i.e., for \(V>V_{\nu}\); and \textit{iii}) the excess current \(I_x(V)\) that appears at \(V\approx V_{\nu}\) \cite{sze_semiconductor}. The I-V characteristic of a typical TD is depicted in Fig.~\ref{fig:I-V characteristic}, where \(I_{\nu}\) is the current right after the tunneling region, while \(V_p\) and \(I_p\) are respectively the peak voltage and current right before entering the tunneling region. 
In this paper, we consider operation in the tunneling region, thus, the cumulative current at the diode \(I(V)\) is described only by \(I_T(V)\), i.e., \(I(V)=I_T(V)\). According to \cite{sze_semiconductor}, this cumulative current is given with respect to the voltage value \(V\) applied at the diode as follows:
\begin{equation}\label{eq:tunneling_current}
    I_T(V)=\frac{I_p}{V_p} V \exp\left(1-\frac{V}{V_p}\right).
\end{equation}

For a given I-V characteristic with fixed \(V_p,\,I_p,\,V_{\nu},\) and \(I_{\nu}\) parameters, the TD operates stably at a single point in the tunneling region, which corresponds to the minimum negative resistance (in absolute terms); we denote this stability point as \(R_{\rm sp}\). This stability results from the fact that the derivative of the resistance with respect to voltage is zero, ensuring robustness to slight biasing voltage variations. To achieve different values for the negative resistance, the diode's parameters need to be altered. For example, the parameters \(V_p\) and \(I_p\) depend on the temperature, doping levels of the p-n junction, effective tunneling mass, and bandgap~\cite[Fig.~68]{BERGER2011176}. Increasing the doping of both n- and p-sides sharpens the I-V characteristic in the tunneling region, resulting in higher amplitude negative resistances \cite{sze_semiconductor, BERGER2011176}. Additionally, tunable changes to the I-V characteristic can be achieved by incorporating a transistor in parallel with the negative resistance circuitry~\cite[Fig.~39]{BERGER2011176}. 

To conveniently describe variations in the I-V characteristic, like the aforementioned, via a compact representation, and consequently analyze the performance of the resulting active RIS devices under different negative resistance values, we adopt the tunneling current model of~\cite{modelling_patent}, according to which, for \(V_0 \in [0.1, 0.5]\) in Volts and \(m \in [1, 3]\), it holds that: 
\begin{equation}
    I_T(V) = \frac{V}{R_0} \exp\left(-\left(\frac{V}{V_0}\right)^m\right),
\end{equation}
where \(R_0\) denotes the Ohmic resistance in the TD's linear region, i.e., for \(V < V_p\)). To this end, the negative resistance \(R(V)\), as a function of the applied voltage, is defined as the inverse of the derivative of the current with respect to voltage:
\begin{equation}\label{eq:Negative_resistance_definition}
\begin{split}
    R(V) & \triangleq \left(\frac{d I_T(V)}{dV}\right)^{-1} \\
    & = R_0 \exp\left(\left(\frac{V}{V_0}\right)^m\right)\left(1-m \left(\frac{V}{V_0}\right)^m\right)^{-1}.
    \end{split}
\end{equation}
As previously described, the stable operating point \(R_{\rm sp}\) of the diode for this model can be obtained by solving \(\frac{d R(V)}{d V} = 0\), which yields \(V_{r} = (m^{-1} + 1)^{1/m} V_0\). Then, by substituting \(V_{ r}\) into \eqref{eq:Negative_resistance_definition} results in the \(R_{\rm sp}\) expression:
\begin{equation}\label{eq:R_n}
  R_{\rm sp}\triangleq R(V_{r})= - R_0 m^{-1}\exp\left(\frac{m+1}{m}\right) .
\end{equation}

For the remainder of the paper, the feasible TD-based negative resistance values \(R_n\) for the considered active RIS unit elements are computed using \eqref{eq:R_n}, i.e., \(R_n = R_{\rm sp}\), with \(m \in [1, 3]\) and a given \(R_0\) (e.g., \(R_0 = 1.5 \, \Omega\) \cite[Table~2]{kriplani2011modelling}).

\subsection{Power Consumption Modeling}
The power consumption $P_n$ required to configure the previously modeled TD-based negative resistance of each \(n\)-th active RIS unit element can be computed as follows. It is first noted that, when \(R_n \geq 0\), the whole circuit is passive, resulting in zero power consumption, i.e., \(P_n = 0\). For \(R_n < 0\), the consumed power can be obtained using Section~\ref{tunnel_diode_model} as follows:  
\begin{eqnarray}\label{eq:Pn_m}
P(R_n) \triangleq I_T(V_{ r})V_{ r} = \frac{V_0^2}{R_0}(m^{-1} + 1)^{2/m},  
\end{eqnarray}
which is a function of \(R_n\) due to \eqref{eq:R_n}. To express this power consumption solely as a function of \(R_n\), which will be needed later on in our active RIS design optimization framework, we solve for \(m\) in \eqref{eq:R_n} using the principal Lambert function \(\mathcal{W}_0(\cdot)\)~\cite{Lambert_function}, yielding the relationship:
\begin{equation}\label{eq:m_wrt_R_n}
     m = \frac{1}{\mathcal{W}_0\left(-\frac{R_n}{R_0 e}\right)}.
\end{equation}
Substituting this result into \eqref{eq:Pn_m}, $P(R_n)$ can re-expressed as:
\begin{equation}\label{eq:Pn}
    P(R_n) = \frac{V_0^2}{R_0} \left(\mathcal{W}_0\left(-\frac{R_n}{R_0 e}\right) + 1\right)^{2\mathcal{W}_0\left(-\frac{R_n}{R_0 e}\right)}.
\end{equation}

It can be seen from \eqref{eq:Pn_m} that the power consumption of each active RIS element depends on the tunneling current model parameters $V_0$, $R_0$, and $m$. However, recall from~\eqref{eq:R_n} that we assumed $R_0$ being fixed, since it models the ohmic resistance before the tunneling region, hence, the variability of the negative resistance $R_n$ depends only on \(m \in [1, 3]\). Thus, it can be concluded that, by also fixing $V_0$ to its minimum value of $0.1$ Volts, $P(R_n)$ is minimized with respect to $V_0$; this value will be used throughout this paper. Consequently, $P(R_n)$ as well as \(R_n\) will depend solely on the choice of \(m\). 


\subsection{Proposed Phase-Amplitude Dependence Model}\label{sec:amplitude_phase}
We now map the amplitude \(\alpha_n\) and phase \(\varphi_n\) of each \(n\)-th active RIS unit element to the corresponding \(R_n\) and \(C_n\) values. To this end, we express each $n$-th reflection coefficient in~\eqref{eq:ref} as \([\boldsymbol{\gamma}]_n(\alpha_n,\phi_n) = \alpha_n e^{\jmath \varphi_n}\), and derive the amplitude boundaries, for a given \(\varphi_n\), in terms of \(R_n\) and \(C_n\) as follows:
\begin{eqnarray}
\alpha_{\rm min}(\varphi_n)&\triangleq&  \sqrt{\frac{Z^{\rm I}_{n,{\rm max}} + (Z_0-Z^{\rm R}_{n,{\rm max}})\tan(\varphi_n)}{Z^{\rm I}_{n,{\rm max}} + (Z_0+Z^{\rm R}_{n,{\rm max}})\tan(\varphi_n) }}, \label{eq:amplitude_min}\\
\alpha_{\rm max}(\varphi_n)&\triangleq& \sqrt{\frac{Z^{\rm I}_{n,{\rm min}} + (Z_0-Z^{\rm R}_{n,{\rm min}})\tan(\varphi_n)}{Z^{\rm I}_{n,{\rm min}} + (Z_0+Z^{\rm R}_{n,{\rm min}})\tan(\varphi_n) }}, \label{eq:amplitude_max}
\end{eqnarray}
%
where \(Z_{n,{\rm min}} \triangleq Z_n(C_n(R_{\max},\varphi_n), R_{\rm max})\) and \(Z_{n,{\rm max}} \triangleq Z_n(C_n(R_{\min},\varphi_n), R_{\rm min})\) with \(R_{\rm min}\) and \(R_{\rm max}\) denoting the lowest and highest resistance values, respectively. The superscripts \({\rm R}\) and \({\rm I}\) indicate the real and imaginary parts of the complex number involved. Furthermore, the term \(C_n(R_n, \varphi_n)\) represents the required capacitance for each \(n\)-th RIS unit element to achieve a specific phase shift \(\varphi_n\) for a given \(R_n\) value. To acquire \(C_n(R_n,\varphi_n)\), we substitute \eqref{eq:Z_n} into \eqref{eq:ref}, yielding the equation in~\eqref{eq:solve_wrt_Cn}. 
\begin{figure*}[t]
\vspace{0.1 cm}
\begin{equation}\label{eq:solve_wrt_Cn}
        \frac{2Z_0(\omega^3L_1L_2(L_1+L_2)-2\omega L_1L_2/C_n(R_n,\varphi_n)-\omega L_1^2/C_n(R_n,\varphi_n)+L_1/(\omega C_n^2(R_n,\varphi_n))+\omega L_1R_n^2)}{(\omega L_1R_n)^2+(L_1/C_n(R_n,\varphi_n)-\omega^2L_1L_2)^2-Z_0^2(\left(\omega(L_1+L_2)-1/(\omega C_n(R_n,\varphi_n))\right)^2+R_n^2)}=\tan(\varphi_n)
\end{equation}
\hrulefill
\end{figure*}
It can be easily concluded that this equation has the following two solutions for \(C_n(R_n, \varphi_n)\):  
\begin{equation}\label{eq:Cn_diakrinousa}
      C_n(R_n,\varphi_n) \! = \! \frac{-b(\varphi_n)\! \pm \! \sqrt{b^2(\varphi_n) \! - \! 4a(R_n,\varphi_n)c(\varphi_n)}}{2 a(R_n,\varphi_n)},\!
\end{equation}
where we have defined the coefficients:
\begin{align*}
        a(R_n,\varphi_n) \triangleq & 2 Z_0 \omega^3 L_1 L_2 (L_1 + L_2) + 2 Z_0 \omega L_1 R_n^2  \\
        & - \omega^2 L_1^2 R_n^2 \tan(\varphi_n) + 
Z_0^2 \omega^2 (L_1 + L_2)^2 \tan(\varphi_n), \\
& - (\omega^2 L_1 L_2)^2 \tan(\varphi_n) + Z_0^2 R_n^2 \tan(\varphi_n)\\
b(\varphi_n) \triangleq & 2 L_1^2 L_2 \omega^2 \tan(\varphi_n) - 4 Z_0 \omega L_1 L_2 \\
&- 2 Z_0 \omega L_1^2 - 2 Z_0^2 (L_1 + L_2) \tan(\varphi_n),\\
c(\varphi_n) \triangleq & 2 Z_0 \frac{L_1}{\omega} - L_1^2 \tan(\varphi_n) + \frac{Z_0^2}{\omega^2} \tan(\varphi_n).
\end{align*}
However, only one of these solutions provides the desired phase shift \(\varphi_n\), while the other corresponds to \(\varphi_n + \kappa \pi\), with \(\kappa = \pm 1\), due to the periodicity of the \(\tan(\cdot)\) function. Moreover, for certain pairs \((R_n, \varphi_n)\), the square root in \eqref{eq:Cn_diakrinousa} may be undefined, indicating the inexistence of feasible solutions for the tunable capacitance. 

To deal with the latter issue, we now determine the range of \(R_n\) values that permits a feasible \(C_n(R_n, \varphi_n)\) for a given \(\varphi_n\) value. To this end, we introduce the auxiliary constants \(a_{i}\), \(b_{j}\), and \(c_{j}\), where \(i=0,1,2,3\) and \(j=0,1\), such that the following equations hold:
\begin{align*}
        a(R_n,\varphi_n) = \, & a_{3}R_n^2\tan(\varphi_n) + a_{2}\tan(\varphi_n) + a_{1}R_n^2 + a_{0},\\
        b(\varphi_n) = \,& b_{1} \tan(\varphi_n) + b_{0},\\
        c(\varphi_n) =\, & c_{1}\tan(\varphi_n) + c_{0}.
\end{align*}
Using these definitions, we can derive the interval for \(R_n\) as:
\begin{equation}\label{eq:R_interval_for_phi}
   | R_n| \! \leq \! \sqrt{\left| \frac{b^2(\varphi_n) \! - \! 4 (a_2\tan(\varphi_n)\! + \! a_0)c(\varphi_n)}{4 c(\varphi_n) (a_3 \tan(\varphi_n) + a_1)} \right|} \! \triangleq \! \mathcal{F}(\varphi_n),
\end{equation}
where function \(\mathcal{F}(x)\), with \(x \in [0, 2\pi]\), specifies the range of \(R_n\) values that yield a feasible \(C_n(R_n, x)\) for a given \(x = \varphi_n\). This range must remain non-zero \(\forall x\), otherwise, certain phase shift values would only be achievable when \(R_n = 0\), thereby, preventing effective control over \(R_n\). To overcome this, \(\mathcal{F}(x)\) should have zero real roots, or equivalently, it should hold that:

\begin{equation}\label{eq:Rn_diakrinousa}
    (2 b_1b_0 - 4 a_2c_0-4a_0c_1)^2 - 4(b_1^2-4a_2c_1)(b_0^2-4a_0c_0)  \leq  0,
\end{equation}
Clearly, this inequality depends solely on the constant system parameters \(L_1\), \(L_2\), \(Z_0,\) and \(\omega\), which should be selected to ensure that this condition is always satisfied. Moreover, the \(R_{\max}\) and \(R_{\min}\) values in~\eqref{eq:amplitude_min} and \eqref{eq:amplitude_max} must comply with inequality~\eqref{eq:R_interval_for_phi}, implying that they should both be functions of \(\varphi_n\); we represent them as \(R_{\max}(\varphi_n)\) and \(R_{\min}(\varphi_n)\). Thus, the corresponding impedance limits in~\eqref{eq:amplitude_min} and \eqref{eq:amplitude_max} are expressed as \(Z_{n,{\rm min}} = Z_n(C_n(R_{\max}(\varphi_n), \varphi_n), R_{\max}(\varphi_n))\) and \(Z_{n,{\rm max}} = Z_n(C_n(R_{\min}(\varphi_n), \varphi_n), R_{\min}(\varphi_n))\).

In conclusion, \(R_{\max}(\varphi_n)\) and \(R_{\min}(\varphi_n)\) can be determined using \eqref{eq:R_interval_for_phi}, whereas \(C_n(R_{\max}(\varphi_n), \varphi_n)\) and \(C_n(R_{\min}(\varphi_n), \varphi_n)\) can be computed via~\eqref{eq:Cn_diakrinousa}. These values enable the calculation, using \eqref{eq:amplitude_min} and \eqref{eq:amplitude_max}, of the amplitude boundaries for the active RIS unit elements as functions of the imposed phase shifts.

\subsection{Reflection Vector Representation}\label{subsec:RIS reflection vector representation}
\begin{figure}[!t]
 \includegraphics[width=\columnwidth]{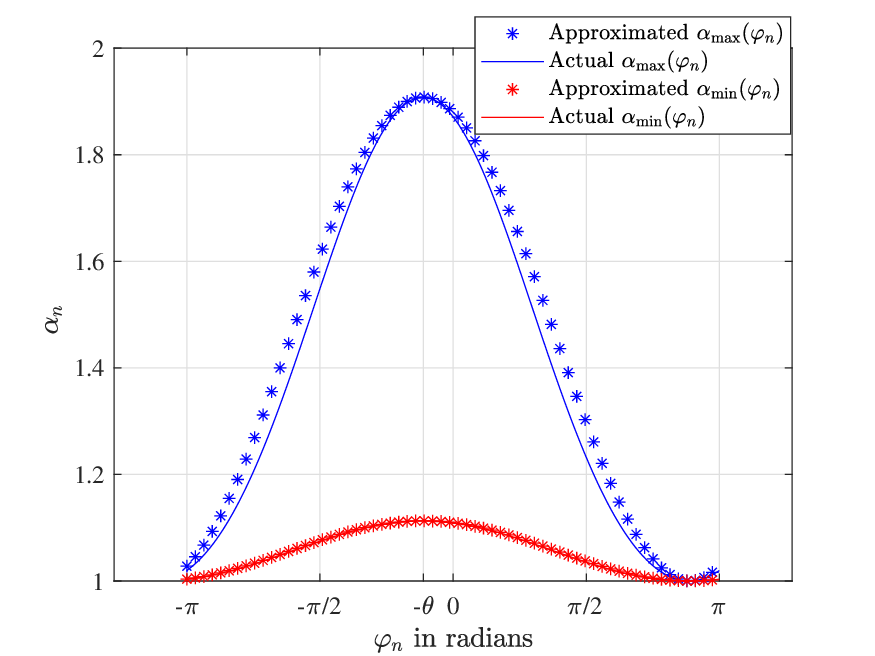}
 \caption{The upper, \(\alpha_{\max}(\varphi_n)\), and lower, \(\alpha_{\min}(\varphi_n)\), bounds for the amplitude for each \(n\)-th RIS active unit element as a function of the tunable phase value \(\varphi_n\), according to the proposed active transmission line circuit model. The derived analytical formulas in Section~\ref{sec:amplitude_phase} are compared with the respective approximations~\eqref{eq:amplitude_max_approx} and \eqref{eq:amplitude_min_approx} in Section~\ref{subsec:RIS reflection vector representation}, considering $L_1$ = 4.5 nH, $L_2$ = 0.7 nH, $Z_0 = 377 \, \Omega$, $\omega = 2 \pi \times 2.4$ GHz, \(R_0 = 0.5\,\Omega\), \(R_n\equiv R_{\rm sp} \in [-3.7,-0.64]\)~$\Omega$ in~\eqref{eq:R_n}, and \(C_n \in [0.85,6.25]\) pF.}
 \label{fig:phase_amplitude_approximation}
 \end{figure}

In this section, we use a linear steepness approximation (via setting \(k=1\) in \cite[eq. (5)]{Abeywickrama_2020}) to  derive an approximate closed-form representation for \(\alpha_{\rm max}(\varphi_n)\) and \(\alpha_{\rm min}(\varphi_n)\) solely with respect to the applied phase value \(\varphi_n\), thereby, eliminating their dependence on the circuit parameters. Our approach extends the passive transmission line circuit model of~\cite{Abeywickrama_2020} to our active RIS unit elements. To this end, we approximate the previously derived exact amplitude boundaries as follows:
\begin{align}
\alpha_{\rm min}(\varphi) \cong & \frac{\delta_{\max}-\delta_{\min}}{2}\left( \cos(\varphi + \theta) + 1 \right) + \delta_{\min}, \label{eq:amplitude_min_approx}\\
\alpha_{\rm max}(\varphi) \cong & \frac{\beta_{\max}-\beta_{\min}}{2}\left( \cos(\varphi + \theta) + 1 \right) + \beta_{\min}, \label{eq:amplitude_max_approx}
\end{align}
where we have introduced the following fitting variables to capture the effect of the negative resistance on the reflection coefficient's amplitude: \(\delta_{\max} \triangleq \max_{\varphi} \alpha_{\min}(\varphi),\,\delta_{\min} \triangleq \min_{\varphi} \alpha_{\min}(\varphi),\,\beta_{\max} \triangleq \max_{\varphi} \alpha_{\max}(\varphi),\) and \(\beta_{\min} \triangleq \min_{\varphi} \alpha_{\max}(\varphi)\), while \(\theta\) is equal to minus the phase shift where \(\delta_{\max}\) and \(\beta_{\max}\) occur.
The accuracy of the approximations~\eqref{eq:amplitude_min_approx} and~\eqref{eq:amplitude_max_approx}, in comparison with their previously derived respective exact expressions~\eqref{eq:amplitude_min} and~\eqref{eq:amplitude_max}, is numerically investigated in Fig.~\ref{fig:phase_amplitude_approximation} for an example setting of the RIS element's circuit parameters. As it can be observed, both approximations are sufficiently tight, with the approximation for \(\alpha_{\min}(\varphi_n)\) being slightly tighter than that for \(\alpha_{\max}(\varphi_n)\).

Capitalizing on~\eqref{eq:amplitude_min} and~\eqref{eq:amplitude_max}, the reconfigurable amplitude \(\alpha_n\) for each \(n\)-th active RIS unit element with respect to the phase \(\varphi_n\), using the normalized variable \(\bar{\alpha}_n\in[0,1]\), can be expressed as follows:
\begin{equation}\label{eq:amplitude_wrt_phase}
    \alpha_n = \alpha_{\min}(\varphi_n) + \bar{\alpha}_n (\alpha_{\max}(\varphi_n) -\alpha_{\min}(\varphi_n) ).
\end{equation}
This formulation enables us to augment the RIS reflection amplification coefficient expression $[\boldsymbol{\gamma}]_n = \alpha_n e^{\jmath \varphi_n}$, with $\alpha_n \in [\alpha_{\min}(\varphi_n),\alpha_{\max}(\varphi_n)]$, through the following representation:
\begin{equation}\label{eq:temp}
    [\boldsymbol{\gamma}]_n \!\!=\!\! \left(\alpha_{\min}(\varphi_n) \! + \! \bar{\alpha}_n (\alpha_{\max}(\varphi_n) -\alpha_{\min}(\varphi_n) \right)e^{\jmath \varphi_n}.
\end{equation} 
Finally, by invoking the approximate expressions \eqref{eq:amplitude_min_approx} and \eqref{eq:amplitude_max_approx} for \(\alpha_{\min}(\cdot)\) and \(\alpha_{\max}(\cdot)\), respectively, we can formulate the reflection amplification coefficient vector of our \(N\)-element TD-based active RIS as follows:

\begin{align}\label{eq:RIS_reflection coefficient vector}
    \boldsymbol{\gamma} \cong & \underbrace{\diag\left\{0.25 e^{\jmath \boldsymbol{\theta}} \odot \left(\mathbf{y} + \mathbf{x} \odot \boldsymbol{\bar{\alpha}}\right) \right\}}_{\triangleq \mathbf{Z}_2} \boldsymbol{\mathcal{D}}^{\tran}(\boldsymbol{\phi}\otimes\boldsymbol{\phi}) \nonumber\\
    & + \underbrace{\diag\left\{0.5 \mathbf{y} + \boldsymbol{\delta}_{\min} + \left(0.5 \mathbf{x} + \boldsymbol{\beta}_{\min} -\boldsymbol{\delta}_{\min}\right)\odot\boldsymbol{\bar{\alpha}} \right\}}_{\triangleq \mathbf{Z}_1} \boldsymbol{\phi}  \nonumber\\
    & + \underbrace{0.25e^{-\jmath \boldsymbol{\theta}}\odot \left(\mathbf{y} + \mathbf{x} \odot\boldsymbol{\bar{\alpha}} \right) }_{\triangleq \mathbf{z}} \in \mathbb{C}^{N\times 1}, 
\end{align}
where we have used the following definitions:
    $\boldsymbol{\phi} \triangleq  [e^{\jmath \varphi_1}, e^{\jmath \varphi_2}, \ldots, e^{\jmath \varphi_N}]^{\tran} \in \mathcal{S}^N$, with $\mathcal{S}\triangleq \{e^{\jmath \varphi}|\varphi \in [0,2\pi]\}$, $    \boldsymbol{\bar{\alpha}} \triangleq [\bar{\alpha}_1,\bar{\alpha}_2,\ldots, \bar{\alpha}_N] \in [0,1]^N$, $\mathbf{x} \triangleq \boldsymbol{\beta}_{\max}-\boldsymbol{\beta}_{\min}-(\boldsymbol{\delta}_{\max}-\boldsymbol{\delta}_{\min})\in \mathbb{R}^{N}$, $\mathbf{y} \triangleq \boldsymbol{\delta}_{\max} -\boldsymbol{\delta}_{\min}\in \mathbb{R}^{N}$, and  $\boldsymbol{\mathcal{D}}\triangleq\left[\mathbf{D}_1\,\mathbf{D}_2\,\cdots\,\mathbf{D}_N\right]^{\rm T}\in\mathbb{B}^{N^2\times N}$, with $\mathbf{D}_i$ being a square-zero matrix with only its $i$-th principal diagonal element being $1$, and $\mathbb{B}\triangleq\{0,1\}$. Additionally, \(\boldsymbol{\delta}_{\min},\,\boldsymbol{\delta}_{\max},\,\boldsymbol{\beta}_{\min},\,\boldsymbol{\beta}_{\max},\) and \(\boldsymbol{\theta}\) are all \(N\)-element vectors containing the fitting parameter values for each active RIS unit element. In most practical cases, each of these vectors will contain the same values for all RIS elements, since identical unit cells are typically used for a single RIS. This implies, for example, that \([\boldsymbol{\delta}_{\max}]_n=\delta_{\max}\,\forall\,n=1,\ldots,N\); similarly will hold for the rest of the vectors.

The derived approximate expression in~\eqref{eq:RIS_reflection coefficient vector} for the  RIS reflection amplification vector includes the tunable phase vector~\(\boldsymbol{\phi}\) and the tunable vector \(\boldsymbol{\bar{\alpha}}\) with \(\bar{\alpha}_n\in[0,1]\) $\forall n$, controlling each $[\boldsymbol{\gamma}]_n$'s amplitude as shown in~\eqref{eq:temp}, instead of the circuit variables \(R_n\) and \(C_n\) for each \(n\)-th active unit element. Most importantly, the phase-amplitude dependence of the reflection coefficient is inherently incorporated via straightforward linear algebra operations. This features facilitates the application of convex relaxation and gradient descent/ascent algorithms for active RIS optimization, as will be shown later on, which would, otherwise, be impossible due to the intractable formulation of \(\boldsymbol{\gamma}\) with respect to the circuit parameters.

\begin{remark}
\label{remark: active and passive unit cells}
The reflection vector representation in~\eqref{eq:RIS_reflection coefficient vector} for our TD-based active RIS can also incorporate the case where some unit elements possess amplifying capabilities, while others being passive. Only phase reconfiguration can be applied for the latter elements, and their circuits' Ohmic resistances will be positive. For these elements, it will hold for their amplitude \(\alpha_{\rm pass}(\varphi)\) that: \(\alpha_{\min}(\varphi)=\alpha_{\max}(\varphi)=\alpha_{\rm pass}(\varphi)\). Hence, to account for passive elements in \eqref{eq:RIS_reflection coefficient vector}, the vectors \(\mathbf{x},\,\mathbf{y},\,\boldsymbol{\beta}_{\min},\, \boldsymbol{\delta}_{\min}\), and \(\boldsymbol{\theta}\) would have two sets of values, one for the active and the other for passive elements. It is noted that for each \(i\)-th passive element, it holds that \([\boldsymbol{\beta}_{\min}]_i=[\boldsymbol{\delta}_{\min}]_i\) and \([\boldsymbol{\beta}_{\max}]_i=[\boldsymbol{\delta}_{\max}]_i\), yielding \([\mathbf{x}]_i=0\). Consequently, for these elements, the vector \(\boldsymbol{\bar{\alpha}}\) impacting the tunable amplitude parameter becomes nulled, as expected.
\end{remark}


\section{System Model and Problem Formulation} \label{sec:System Model and Problem Formulation}
Consider an $M_{\rm T}$-antenna TX wishing to communicate with an $M_{\rm R}$-antenna RX. The TX transmits \(d \leq \min\{M_{\rm T}, M_{\rm R}\}\) data streams, represented by \(\mathbf{s} \in \mathbb{C}^{d \times 1}\) (in practice, chosen from a discrete constellation set), using spatial precoding via the matrix \(\mathbf{V} \in \mathbb{C}^{M_{\rm T} \times d}\), such that \({\rm Tr}\{\mathbf{V}^H\mathbf{V}\} \leq P_{ T}\) with \(P_{ T}\) being the total TX power budget. This communication is assisted by an RIS with \(N\) active unit elements. The complex-valued baseband received signal at the $M_{\rm R}$ RX antenna elements can be mathematically expressed as \cite[eq.~(1)]{George_RIS_TWC2019} with the addition of the noise generated by the RIS's active components, i.e.:
\begin{align}\label{eq:system_model_antennas_streams}
  \mathbf{y} &\triangleq \underbrace{\left(\mathbf{H}_{\rm d}+\mathbf{H}_{2}\mathbf{\Gamma}\mathbf{H}_{1}\right)}_{\triangleq\tilde{\mathbf{H}}}
  \mathbf{V}\mathbf{s}+\mathbf{H}_2\boldsymbol{\Gamma} \mathbf{n}_s + \mathbf{n}_r,
\end{align}
where \(\mathbf{H}_{\rm d} \in \mathbb{C}^{M_{\rm R} \times M_{\rm T}}\) is the direct RX-TX channel, \(\mathbf{H}_{1} \in \mathbb{C}^{N \times M_{\rm T}}\) is the RIS-TX channel, and \(\mathbf{H}_{2} \in \mathbb{C}^{M_{\rm R} \times N}\) is the RX-RIS channel. The diagonal matrix \(\mathbf{\Gamma} \triangleq \mathrm{diag}\{\boldsymbol{\gamma}\} \in \mathbb{C}^{N \times N}\) models the RIS reflection amplification coefficients, where \(\boldsymbol{\gamma} = \boldsymbol{\alpha} \odot \boldsymbol{\phi}\) with \(\boldsymbol{\alpha} \triangleq [\alpha_1, \ldots, \alpha_N]^{\rm T}\), and \(\alpha_n \in [\alpha_{\rm min}(\varphi_n), \alpha_{\rm max}(\varphi_n)]\). The vector terms \(\mathbf{n}_r \sim \mathcal{CN}(\mathbf{0}_{M_{\rm R} \times 1}, F_r \sigma^2 \mathbf{I}_{M_{\rm R}})\) and \(\mathbf{n}_s \sim \mathcal{CN}(\mathbf{0}_{N \times 1}, F_s \sigma^2 \mathbf{I}_N)\) represent the noises generated at the RX and RIS, respectively, with \(F_r\) and \(F_s\) being the noise figures quantifying the SNR degradation due to the active electronics components at the RX and RIS, respectively~\cite{6047578}. At the RX, a combining matrix \(\mathbf{W} \in \mathbb{C}^{M_{\rm R} \times d}\) is applied in baseband to the received signal, yielding the detection for $\mathbf{s}$: \(\mathbf{\hat{s}} = \mathbf{W}^H\mathbf{y} \in \mathbb{C}^d\).

Given the availability of the channels \(\mathbf{H}_{\rm d}\), \(\mathbf{H}_1\), and \(\mathbf{H}_2\)~\cite{Jian_RIS_survey} to formulate~\eqref{eq:system_model_antennas_streams}, we focus on the joint design of the TX precoder \(\mathbf{V}\), the RX combiner \(\mathbf{W}\), and the RIS phase-amplitude configuration vector \(\boldsymbol{\gamma}\) to maximize the achievable end-to-end MIMO rate. By assuming that \(\mathbf{s} \sim \mathcal{CN}(\mathbf{0}_{d \times 1}, \mathbf{I}_d)\) (i.e., independent Gaussian symbols), the instantaneous spectral efficiency can be expressed as~\cite[eq. (6.61)]{heath}:

\begin{align}\label{eq:R_RX_w_noise}
 &\mathcal{R}(\mathbf{V},\mathbf{W},\boldsymbol{\gamma}) \triangleq \sum_{i=1}^d \log_2 \Biggl(1 + \\
 &\frac{\norm{\mathbf{w}_i^H\tilde{\mathbf{H}}\mathbf{v}_i}_{\rm F}^2}{\sum \limits_{j=1, j \neq i}^d \norm{\mathbf{w}_i^H\tilde{\mathbf{H}}\mathbf{v}_j}_{\rm F}^2 + \sigma^2 F_s \norm{\mathbf{w}_i\mathbf{H}_2\mathbf{\Gamma}}_{\rm F}^2 + \sigma^2 F_r \norm{\mathbf{w}_i}_{\rm F}^2}\Biggr). \nonumber
\end{align}
Our design objective can be thus mathematically formulated as:
\begin{align}
\mathcal{OP}:&\max_{\mathbf{V},\mathbf{W},\boldsymbol{\gamma}} \mathcal{R}(\mathbf{V},\mathbf{W},\boldsymbol{\gamma}) \nonumber \\
&\,\,\,\,{\rm s.t.} \,\, {\rm Tr}\{\mathbf{V}^H\mathbf{V}\} \leq P_{ T}, \,\, \sum_{n=1}^N P(R_n) \leq P_{\rm RIS},\label{OP_constraints_1} \\
&\,\,\,\,0 \leq \varphi_n \leq 2\pi, \,\, \alpha_n \in [\alpha_{\rm min}(\varphi_n), \alpha_{\rm max}(\varphi_n)] \,\, \forall n, \label{OP_constraints_2}
\end{align}
where \(P_{\rm RIS}\) denotes the total available power at the active RIS.

\section{Proposed Joint MIMO and Active RIS Design}\label{sec:solution_methodologies}
In this section, we present two approaches for designing $\mathbf{V}$, $\mathbf{W}$, and $\boldsymbol{\gamma}$ that solve \(\mathcal{OP}\). The first approach is based on AO, decomposing the main rate maximization problem into tractable subproblems, which are solved sequentially in an iterative manner. Relying on high SNR approximations and assuming negligible noise from the active RIS, in comparison with the RX thermal noise contribution, we also devise an one-step optimization approach that can also serve as an initialization joint MIMO and active RIS design for the AO-based algorithm.

\subsection{Iterative Solution}\label{subsec: Iterative optimization methodology}
We now solve \(\mathcal{OP}\) via an AO approach, i.e., independently for each design parameter keeping the others fixed, and then iterating over those individual solutions till convergence. 

\subsubsection{RX Combiner Optimization}
We capitalize on the optimality of Linear Minimum Mean Squared Error (LMMSE) reception with respect to the spectral efficiency~\cite{heath} (this is also visible from \eqref{eq:R_RX_w_noise}'s structure). To this end, each column of \(\mathbf{W}\), defined as \(\mathbf{w}_i\), is obtained from the solution of the generalized eigenvalue problem \(\frac{\mathbf{w}_i^{\herm}\mathbf{N}_i\mathbf{w}_i}{\mathbf{w}_i^{\herm}\mathbf{F}_i\mathbf{w}_i}\), where \(\mathbf{N}_i\triangleq\mathbf{\tilde{H}v}_i\mathbf{v}_i^{\herm}\mathbf{\tilde{H}}^{\herm}\) and \(\mathbf{F}_i\triangleq \sum \limits_{j=1,\,j\neq i}^{d}\mathbf{\tilde{H}}\mathbf{v}_j\mathbf{v}_j^{\herm}\mathbf{\tilde{H}}^{\herm}+\sigma^2 F_s \mathbf{H}_2\mathbf{\Gamma}\mathbf{\Gamma}^{\herm}\mathbf{H}_2^{\herm}+\sigma^2 F_r \mathbf{I}_{M_{\rm R}} \) with \(\mathbf{v}_i\)'s  (\(i=1,\ldots,d\)) indicating the columns of \(\mathbf{V}\). Following this solution and using the definition \(\mathbf{F} \triangleq \mathbf{\tilde{H}}\mathbf{V}\mathbf{V}^{\herm}\mathbf{\tilde{H}}^{\herm} \! +\! \sigma^2 F_s\mathbf{H}_2\mathbf{\Gamma}\mathbf{\Gamma}^{\herm}\mathbf{H}_2^{\herm}\! +\! \sigma^2 F_r\mathbf{I}_{M_{\rm R}} \) resulting from $\mathbf{F}_i$'s, the scaling invariant LMMSE-based RX combining matrix is obtained as:
 \begin{equation}\label{eq:LMMSE_Receiver}
     \mathbf{W_{\rm opt}}=\mathbf{F}^{-1}\mathbf{\tilde{H}V}.
 \end{equation}
The resulting Signal to Interference plus Noise Ratio (SINR) per stream, after some algebraic manipulations \cite[Section~6.4.2]{heath}, can be derived as \({\rm SINR}_i \triangleq\mathbf{v}_i^{\herm}\mathbf{\tilde{H}}^{\herm}\mathbf{F }_i^{-1}\mathbf{\tilde{H}v}_i\), yielding the following expression for the achievable rate:
\begin{equation}\label{eq:Rate_linear_receiver}
    \mathcal{R}\left(\mathbf{V},\boldsymbol{\gamma}\right) =  \sum_{i=1}^{d} \log_2\left(1+ \mathbf{v}_i^{\herm}\mathbf{\tilde{H}}^{\herm}\mathbf{F }_i^{-1}\mathbf{\tilde{H}v}_i\right).
\end{equation}

\subsubsection{TX Precoder Optimization}
The previous rate expression is non-convex with respect to the precoding matrix \(\mathbf{V}\) due to the inverse term appearing inside the logarithmic function. To derive a more tractable expression, we first get the ratios outside the logarithm, and then decouple them~\cite{matrix_transforms_optimization}. For this, we employ the Lagrangian dual and the matrix quadratic transforms following~\cite[Theorems 2 and 3]{matrix_transforms_optimization} to obtain two reformulation for~\eqref{eq:Rate_linear_receiver}. To this end, we introduce the auxiliary variables \(\boldsymbol{\sigma}\triangleq [\sigma_1,\dots,\sigma_d]^{\rm T}\, \in \mathbb{R}^{d\times 1}\) and \(\mathbf{Y}\triangleq [\mathbf{y}_1,\dots,\mathbf{y}_d]\, \in \mathbb{C}^{M_{\rm R} \times d}\) to equivalently express \(\mathcal{OP}\) as follows:
\begin{align*}
\overline{\mathcal{OP}}\!:&\max_{\mathbf{V,Y},\boldsymbol{\sigma},\boldsymbol{\gamma}} \mathcal{R}\left(\mathbf{V},\mathbf{Y},\boldsymbol{\sigma},\boldsymbol{\gamma}\right) \!=\! \sum_{i=1}^{d}\!\Big(\!\log_2(1\!+\!\sigma_i)\!-\!\sigma_i\!+\!(1\!+\!\sigma_i) \nonumber\\
    & \hspace{3.9cm} \times\! \Big(\!2\Re[\mathbf{v}_i^{\herm}\mathbf{\tilde{H}}^{\herm}\mathbf{y}_i]\!-\!\mathbf{y}_i^{\herm}\mathbf{F}\mathbf{y}_i\Big)\!\Big) \nonumber\\
    & {\hspace{0.4cm}\text{s.t. constraints in \eqref{OP_constraints_1} and \eqref{OP_constraints_2}},}
\end{align*}
which includes a concave objective with respect to both \(\boldsymbol{\sigma}\) and \(\mathbf{Y}\). In fact, by keeping the other variables fixed, the optimal \(\mathbf{y}_i\)'s and \(\sigma_i\)'s are obtained as \(\mathbf{y}_{i,{\rm opt}}=\mathbf{F}^{-1}\mathbf{\tilde{H}}\mathbf{v}_i\) and \(\sigma_{i,{\rm opt}}={\rm SINR}_i\), respectively. Thereon, to solve for \(\mathbf{V}\), we reformulate $\overline{\mathcal{OP}}$ in a matrix form, yielding the following optimization problem:
\begin{align*}
\mathcal{OP}_{\mathbf{V}}:\,\,&\max_{\mathbf{V}}  f(\mathbf{V}) \triangleq  \trace\{\mathbf{\Sigma}\mathbf{V}^{\herm}\mathbf{\tilde{H}}^{\herm}\mathbf{Y}\} + \trace\{\mathbf{\Sigma}\mathbf{Y}^{\herm}\mathbf{\tilde{H}}\mathbf{V}\} \nonumber\\
&\hspace{2cm} - \trace\{\mathbf{\Sigma}\mathbf{Y}^{\herm}\mathbf{F}\mathbf{Y}\}
\nonumber\\
&\,\,{\rm s.t.}\,\,{\rm Tr}\{\mathbf{V}^H\mathbf{V}\}\leq P_{ T},
\end{align*}
where \(\mathbf{\Sigma}\triangleq {\rm diag}(\boldsymbol{\sigma}+\mathbf{1}_{d\times 1})\). It can be observed that \(\mathcal{OP}_{\mathbf{V}}\) is concave with respect to \(\mathbf{V}\), hence, to acquire the optimal TX precoder, we apply the following KKT conditions:
\begin{itemize}
    \item Let \(\lambda_{\rm opt}\geq 0\) denote the optimal KKT multiplier.
    \item It should hold from the complementary slackness that: \(\lambda_{\rm opt}\left(\trace\{\mathbf{V}_{\rm opt}\mathbf{V}_{\rm opt}^{\herm}\}  -  P_T\right) = 0\).
    \item Then, \(\mathbf{V}_{\rm opt} =\left(\mathbf{\tilde{H}}^{\herm}\mathbf{Y\Sigma}\mathbf{Y}^{\herm}\mathbf{\tilde{H}} + \lambda_{\rm opt}\mathbf{I}_{M_T}\right)^{-1}\mathbf{\tilde{H}}^{\herm}\mathbf{Y\Sigma}\).
\end{itemize}

We now define the auxiliary matrices \(\mathbf{Z} \triangleq \mathbf{\tilde{H}}^{\herm}\mathbf{Y\Sigma}\) and \(\mathbf{K} \triangleq \mathbf{\tilde{H}}^{\herm}\mathbf{Y\Sigma}\mathbf{Y}^{\herm}\mathbf{\tilde{H}}\), where \(\mathbf{K}\) is Hermitian positive semidefinite, thus, it can be expressed via its eigenvalue decomposition as \(\mathbf{K} = \mathbf{U}_{K}\mathbf{\Lambda}_{K}\mathbf{U}^{\herm}_{K}\). Subsequently, the optimal solution \(\mathbf{V}_{\rm opt}\) can be substituted into the complementary slackness condition to determine \(\lambda_{\rm opt}\). Following analogous steps to \cite[Proposition~1]{MSE_matrix}, while adapting them to our system model, \(\lambda_{\rm opt}\) is obtained by solving the following equation:
\begin{equation}\label{eq:complement_slackness_equation}
    \sum_{i=1}^{M_T} \frac{[\mathbf{U}^{\herm}_{K}\mathbf{ZZ}^{\herm}\mathbf{U}_{K}]_{i,i}}{\left([\mathbf{\Lambda}_{K}]_{i,i}+\lambda_{\rm opt}\right)^2}=P_T.
\end{equation}
This equation can be easily solved using any one-dimensional search algorithm, e.g., the bisection method. It is noted that the non-zero singular values of \(\mathbf{V}_{\rm opt}\) determine the number of data streams to be actually transmitted.

\subsubsection{RIS Phase Configuration Optimization} \label{subsec: RIS phase configuration}
The objective function \(f(\cdot)\) in $\mathcal{OP}_{\mathbf{V}}$ depends also on \(\boldsymbol{\phi}\) which is included in $\tilde{\mathbf{H}}$. By using the cyclic-shift property of the trace function and the vectorization properties in \cite[1.11.22-1.11.24]{matrix_analysis}, \(f(\cdot)\) can be reformulated with respect to the RIS amplitude-phase configuration \(\boldsymbol{\gamma}\), yielding the following optimization problem:
\begin{align*}
    \mathcal{OP}_{\boldsymbol{\gamma}}:\max_{\boldsymbol{\gamma}}\,  & 2 {\rm Re}\left[ {\rm vec}\left(\mathbf{Y\Sigma}\mathbf{V}^{\herm}\right)^{\herm}\boldsymbol{\mathcal{H}}\boldsymbol{\gamma} \right] \\
    & - \sigma^2 F_s\boldsymbol{\gamma}^{\herm}\boldsymbol{\mathcal{D}}^{\rm T}\left(\mathbf{I}_{N}\otimes \mathbf{H}_2^{\herm}\mathbf{Y}\mathbf{\Sigma}\mathbf{Y}^{\herm}\mathbf{H}_2\right)\boldsymbol{\mathcal{D}}\boldsymbol{\gamma} \nonumber \\
    & - \boldsymbol{\gamma}^{\herm}\boldsymbol{\mathcal{H}}^{\herm}( \mathbf{V}^{\ast}\mathbf{V}^{\rm T}\otimes \mathbf{Y\Sigma}\mathbf{Y}^{\herm})\boldsymbol{\mathcal{H}}\boldsymbol{\gamma} \nonumber\\
    & -2 {\rm Re}\left[{\rm vec}\left(\mathbf{Y}\mathbf{\Sigma}\mathbf{Y}^{\herm}\mathbf{H}_{\rm d}\mathbf{VV}^{\herm}\right)^{\herm} \boldsymbol{\mathcal{H}}\boldsymbol{\gamma}\right],
    \\
    & {\hspace{-0.6cm}\text{s.t. constraints in \eqref{OP_constraints_2}},}
\end{align*}
where we have used the definition \(\boldsymbol{\mathcal{H}}\triangleq(\mathbf{H}_1^{\tran}\otimes \mathbf{H}_2)\boldsymbol{\mathcal{D}}\in \mathbb{C}^{M_T M_R\times N}\). Hereinafter, we will first solve for the phase configuration vector \(\boldsymbol{\phi}\), and later on, given this solution, the optimum amplitude configuration vector \(\boldsymbol{\alpha}\) will be obtained. 

To efficiently account for the phase-amplitude dependence, we deploy our approximate expression in~\eqref{eq:RIS_reflection coefficient vector} to re-express \(\mathcal{OP}_{\boldsymbol{\gamma}}\) as a function of \(\boldsymbol{\phi}\) for a given \(\boldsymbol{\bar{\alpha}}\) value, as follows:
\begin{align}
        \mathcal{OP}_{\boldsymbol{\phi}}: \, \min_{\boldsymbol{\phi}}\,& g(\boldsymbol{\phi})\triangleq  \mathbf{z}^{\rm H} \mathbf{T} \mathbf{z} + 2{\rm Re}\big[(\boldsymbol{\phi} \otimes \boldsymbol{\phi})^{\rm H} \boldsymbol{\mathcal{D}} \mathbf{Z}_2^{\rm H} \mathbf{T} \mathbf{z} \nonumber\\
        & +\boldsymbol{\phi}^{\rm H} \mathbf{Z}_1^{\rm H} \mathbf{T} \mathbf{z}\big] + (\boldsymbol{\phi} \otimes \boldsymbol{\phi})^{\rm H} \boldsymbol{\mathcal{D}} \mathbf{Z}_2^{\rm H} \mathbf{T} \mathbf{Z}_2 \boldsymbol{\mathcal{D}}^{\rm T} (\boldsymbol{\phi} \otimes \boldsymbol{\phi}) \nonumber\\
        & + 2 {\rm Re} \left[ (\boldsymbol{\phi} \otimes \boldsymbol{\phi})^{\rm H} \boldsymbol{\mathcal{D}} \mathbf{Z}_2^{\rm H} \mathbf{T} \mathbf{Z}_1 \boldsymbol{\phi} \right] 
        + \boldsymbol{\phi}^{\rm H} \mathbf{Z}_1^{\rm H} \mathbf{T} \mathbf{Z}_1 \boldsymbol{\phi} \nonumber \\
        & - 2{\rm Re}\left[\left( \mathbf{Z}_2 \boldsymbol{\mathcal{D}}^{\rm T} (\boldsymbol{\phi} \otimes \boldsymbol{\phi}) + \mathbf{Z}_1 \boldsymbol{\phi} + \mathbf{z} \right)^{\rm H} \mathbf{q}\right]\nonumber\\
        & \hspace{-0.5cm}{\rm s.t.}\, \boldsymbol{\phi}\in \mathcal{S}^N, \label{eq:OP_phi_new_model}
\end{align}
where \(\mathbf{Z}_1\), \(\mathbf{Z}_2,\) and \(\mathbf{z}\) have defined in \eqref{eq:RIS_reflection coefficient vector}, while \(\mathbf{T} \triangleq \sigma^2 F_s\boldsymbol{\mathcal{D}}^{\rm T}\left(\mathbf{I}_{N}\otimes \mathbf{H}_2^{\herm}\mathbf{Y}\mathbf{\Sigma}\mathbf{Y}^{\herm}\mathbf{H}_2\right)\boldsymbol{\mathcal{D}} + \boldsymbol{\mathcal{H}}^{\herm}( \mathbf{V}^{\ast}\mathbf{V}^{\rm T}\otimes \mathbf{Y\Sigma}\mathbf{Y}^{\herm})\boldsymbol{\mathcal{H}}\), and \(\mathbf{q} \triangleq \boldsymbol{\mathcal{H}}^{\herm} {\rm vec}\left(\mathbf{Y\Sigma}\mathbf{V}^{\herm}\right) - \boldsymbol{\mathcal{H}}^{\herm}{\rm vec}\left(\mathbf{Y}\mathbf{\Sigma}\mathbf{Y}^{\herm}\mathbf{H}_{\rm d}\mathbf{VV}^{\herm}\right)\). Due to the existence of the unit modulus constraint as well as \(\mathcal{OP}_{\boldsymbol{\phi}}\)'s non-convex objective, we will resort to Riemannian Manifold Optimization (RMO)~\cite{Riemmanian_optimization_Kostas}. Specifically, \(\mathcal{OP}_{\boldsymbol{\phi}}\) can be formulated as an unconstrained problem on the surface of the complex circle manifold \(\mathcal{S}^N\), thus, manifold-based optimization algorithms can be employed to effectively solve it. Gradient descent/ascent algorithms on Riemannian manifolds consist of the following three main steps at each \(t\)-th algorithmic iteration~\cite{Riemannian_opt_steps}: \textit{i}) computation of the Riemannian gradient \(\nabla^{\rm R}_{\boldsymbol{\phi}} g(\boldsymbol{\phi})\) at the \(t\)-th point \(\nabla^{\rm R}_{\boldsymbol{\phi}} g(\boldsymbol{\phi}|\boldsymbol{\phi}_t)\); \textit{ii}) acquisition of the search direction \(\boldsymbol{\eta}_t\) as well as the step size \(\tau_t\); and \textit{iii}) retraction of the solution to the manifold \(\mathcal{S}^N\). Starting with \textit{i}), we calculate the Euclidean gradient \(\nabla_{\boldsymbol{\phi}}g(\boldsymbol{\phi})\), using $g(\boldsymbol{\phi})$'s definition in $\mathcal{OP}_{\boldsymbol{\phi}}$, as follows:
\begin{align}\label{eq:Euclidean_grad_f_phi}
\nabla_{\boldsymbol{\phi}} g(\boldsymbol{\phi}) = 
 & 2 \bigg[ 
\left( \mathbf{I}_N \otimes \boldsymbol{\phi}^{\rm H} + \boldsymbol{\phi}^{\rm H} \otimes \mathbf{I}_N \right) 
\boldsymbol{\mathcal{D}} \mathbf{Z}_2^{\rm H} \mathbf{T} \mathbf{z} 
+ \mathbf{Z}_1^{\rm H} \mathbf{T} \mathbf{z} \nonumber \\
& + \!\left( \mathbf{I}_N \otimes \boldsymbol{\phi}^{\rm H} \! + \! \boldsymbol{\phi}^{\rm H} \otimes \mathbf{I}_N \right) 
\boldsymbol{\mathcal{D}} \mathbf{Z}_2^{\rm H} \mathbf{T} \mathbf{Z}_2 \boldsymbol{\mathcal{D}}^{\rm T} (\boldsymbol{\phi} \otimes \boldsymbol{\phi}) \nonumber\\
& +\! \mathbf{Z}_1^{\rm H} \mathbf{T} \mathbf{Z}_1 \boldsymbol{\phi} 
\!+ \!\left( \mathbf{I}_N \!\otimes \! \boldsymbol{\phi}^{\rm H} \! +\! \boldsymbol{\phi}^{\rm H}\! \otimes \! \mathbf{I}_N \right) 
\boldsymbol{\mathcal{D}} \mathbf{Z}_2^{\rm H} \mathbf{T} \mathbf{Z}_1 \boldsymbol{\phi} 
 \nonumber\\
& - \left(\mathbf{Z}_2 \boldsymbol{\mathcal{D}}^{\rm T} (\boldsymbol{\phi} \otimes \boldsymbol{\phi}) 
+ \mathbf{Z}_1 \boldsymbol{\phi} + \mathbf{z}\right) \mathbf{q} 
\bigg].
\end{align}
Then, we compute the Riemmanian gradient by mapping \(\nabla_{\boldsymbol{\phi}} g(\boldsymbol{\phi})\) to the tangent space of the optimization manifold \(\mathcal{S}^N\) at \(\boldsymbol{\phi}_t\), which is denoted as \(\mathcal{T}_{\boldsymbol{\phi}_t} \mathcal{S}^N \triangleq \{\mathbf{x} \in \mathcal{S}^N:\, {\rm Re}[\mathbf{x}\odot\boldsymbol{\phi}^*_{t}]=\mathbf{0}_{N\times 1}\}\). This process can be formulated as \cite{Riemannian_opt_steps}:
\begin{equation}\label{eq: Riemmanian grad}
    \nabla_{\boldsymbol{\phi}}^{\rm R} g(\boldsymbol{\phi}|\boldsymbol{\phi}_t) = \nabla_{\boldsymbol{\phi}}g(\boldsymbol{\phi}|\boldsymbol{\phi}_t) - {\rm Re}\left[ \nabla_{\boldsymbol{\phi}}g(\boldsymbol{\phi}|\boldsymbol{\phi}_t) \odot \boldsymbol{\phi}^*\right].
\end{equation}

To implement the aforedescribed RMO steps \textit{ii}) and \textit{iii}), we adopt the approach presented in \cite[Algorithm~1]{Riemmanian_optimization_Kostas}. Therefore, the search direction \(\boldsymbol{\eta}_t\) is determined via the Riemannian gradient, which is refined via the Polak-Ribi\`ere coefficient to incorporate second-order optimality considerations, thus, accelerating convergence.  
The step size \({\tau}_t\) is computed using the Armijo-Goldstein backtracking method, which ensures that, at each $t$-th algorithmic iteration, the objective stays non-increasing. Finally, we have used the unit modulus function \({\rm unit}(\mathbf{x})\triangleq {\mathbf{x}}/{|\mathbf{x}|}\) as the retraction operator. 

\subsubsection{RIS Amplitude Configuration Optimization}
We now turn our attention to the optimization of $\boldsymbol{\alpha}$ in \(\mathcal{OP}_{\boldsymbol{\gamma}}\), given the previously derived \(\boldsymbol{\phi}_{\rm opt}\) solution. To this end, we re-express the objective function and constraints with respect to \(\boldsymbol{\alpha}\), as:
\begin{equation*}
    \begin{split}
        \mathcal{OP}_{\boldsymbol{\alpha}}:\, &\min_{\boldsymbol{\alpha}}  g(\boldsymbol{\alpha})\triangleq\boldsymbol{\alpha}^{\tran}\mathbf{\Phi}^{\herm}\mathbf{T}\mathbf{\Phi}\boldsymbol{\alpha} - 2{\rm Re}\left[\mathbf{q}^{\herm}\mathbf{\Phi}\right]\boldsymbol{\alpha} \\
    & \,\,{\rm s.t.}\, \mathbf{l}\leq \boldsymbol{\alpha} \leq \mathbf{u},\,\,\sum_{n=1}^N P(R_n)\leq P_{\rm RIS},
    \end{split}
\end{equation*}
where \(\mathbf{l}\triangleq[\alpha_{\min,1},\dots,\alpha_{\min,N}]^{\tran}\in\mathbb{R}^{N}\), \(\mathbf{u}\triangleq[\alpha_{\max,1},\dots,\alpha_{\max,N}]^{\tran}\in\mathbb{R}^{N}\), and \(\mathbf{\Phi}\triangleq {\rm diag}\left(\boldsymbol{\phi}\right)\in \mathbb{C}^{N\times N}\). Note that we returned to the initial formulation \(\boldsymbol{\gamma}=\boldsymbol{\alpha}\odot \boldsymbol{\phi}\) for the RIS reflection amplification vector, since \(\boldsymbol{\phi}\) is considered fixed herein and the amplitude boundaries have been already set. To solve \(\mathcal{OP}_{\boldsymbol{\alpha}}\), we need to relate the power consumed at the RIS with all amplitudes \(\alpha_n\)'s of its active unit elements. However, the function \(P(\alpha_n)\) is non-linear and difficult to obtain due to the non-linear dependencies between \(R_n\) and \(\alpha_n\) as well as between \(R_n\) and \(P(R_n)\). In fact, it can be seen from \eqref{eq:Pn} that \(P_n(R_n)\) is strictly decreasing with \(R_n\), and \(\alpha_n\) strictly decreases with \(R_n\) as well. This indicates that \(P(\alpha_n)\) is a strictly increasing function of \(\alpha_n\), thus, to approximately solve \(\mathcal{OP}_{\boldsymbol{\alpha}}\), we propose the following linear fitting for \(P(\alpha_n)\):
\begin{align}\label{eq:linear_fitting}
   P(\alpha_n) \cong y(\alpha_n) = & P_{\min,n} \\
   & +(\alpha_n-\alpha_{\min,n})\frac{P_{\max,n}-P_{\min,n}}{\alpha_{\max,n}-\alpha_{\min,n}}, \nonumber
\end{align}
where we have defined \(P_{\min,n}\triangleq P({R_{\max,n}})\) and \(P_{\max,n}\triangleq P({R_{\min,n}})\); recall from Section~\ref{sec:amplitude_phase} that \(R_{\max,n}\) and \(R_{\min,n}\) depend on the phase value of each \(n\)-th element (\(n\!=\!1,\ldots,N\)).

Since both the objective function and the constraints in \(\mathcal{OP}_{\boldsymbol{\alpha}}\) are convex, this problem can be solved via conventional convex optimization tools. Its solution $\boldsymbol{\alpha}_{\rm opt}$ is then used to acquire the corresponding values \(C_n\) and \(R_n\) for the circuit parameters of the active RIS unit elements. To this end, we use the definition \(X_n\triangleq R_n+\frac{1}{\jmath\omega C_n}\) to re-write \eqref{eq:ref} as follows: 
\begin{equation}\label{eq:ref2}
   [\boldsymbol{\gamma}]_n= \frac{\frac{\jmath\omega L_1(\jmath\omega L_2+X_n)}{\jmath\omega L_1+(j\omega L_2+X_n)} - Z_0}{\frac{\jmath\omega L_1(\jmath\omega L_2+X_n)}{\jmath\omega L_1+(j\omega L_2+X_n)} + Z_0},
\end{equation}
For a given $[\boldsymbol{\gamma}]_n$, we can solve \eqref{eq:ref2} with respect to $X_n$ as:
\begin{equation}\label{eq:Xn}
    X_n \! =\! \frac{\omega \! \left(\left( [\boldsymbol{\gamma}]_n \! + \! 1\right) \! Z_0 \! \left(L_1\! +\! L_2\right)\! + \! \jmath L_1 L_2 \left( [\boldsymbol{\gamma}]_n\! -\! 1\right) \omega \right)}{L_1\left( [\boldsymbol{\gamma}]_n-1\right) \omega+\jmath\left( [\boldsymbol{\gamma}]_n+1\right) Z_0},
\end{equation}
yielding the solutions \(R_n=\Re[X_n]\) and \(C_n=\frac{1}{|\Im[X_n]|\omega}\). To compute the actual RIS power consumption with this solution, we calculate \(\sum_{n=1}^{N}P(R_n)\) via \eqref{eq:Pn}. If the value of this summation is less than \(P_{\rm RIS}\), the process is terminated. Otherwise, we decrease the available \(P_{\rm RIS}\) by the error term \(e \triangleq \sum_{n=1}^{N}\left( y(\alpha_n) - P(R_n)\right)\), and solve again \(\mathcal{OP}_{\boldsymbol{\alpha}}\) for this new value. This approach is repeated until the  \(P_{\rm RIS}\) power constraint is satisfied.

\subsubsection{Overall Algorithm}
Our AO-based iterative algorithm for solving $\mathcal{OP}$ computes sequentially at each $j$-th algorithmic iteration the solutions \(\mathbf{Y}_{\rm opt}^{(j)}\), \(\boldsymbol{\sigma}_{\rm opt}^{(j)}\), \(\mathbf{V}_{\rm opt}^{(j)}\), \(\boldsymbol{\phi}_{\rm opt}^{(j)}\), asnd \(\boldsymbol{\alpha}_{\rm opt}^{(j)}\). This procedure is repeated until rate convergence within a specified threshold \(\epsilon\), or till a maximum number of iterations \(J_{\rm alt}\) is met. The final joint design of \(\mathbf{V}_{\rm opt}\), \(\boldsymbol{\phi}_{\rm opt}\), and \(\boldsymbol{\alpha}_{\rm opt}\), and consequently \(\mathbf{W}_{\rm opt}\) via \eqref{eq:LMMSE_Receiver}, is the one yielding the largest achievable rate across all algorithmic iterations. The overall iterative joint MIMO and active RIS design is summarized in Algorithm~\ref{alg:alternate_optimization}. Therein, we have dropped the subscript ``\({\rm opt}\)'' within the iterations to lighten notation.

\begin{algorithm}[t]
\caption{Iterative Joint MIMO and Active RIS Design}
\label{alg:alternate_optimization}
\textbf{Inputs:} \(\mathbf{H}_{\rm d}\), \(\mathbf{H}_1\), \(\mathbf{H}_2\), \(P_{T}\), \(P_{\rm RIS}\), \(d\), \(\epsilon\), and feasible starting points \(\mathbf{V}^{(0)}\), \(\boldsymbol{\phi}^{(0)}\), and \(\boldsymbol{\alpha}^{(0)}\). \\
\textbf{Outputs:} \(\mathbf{V}_{\rm opt}\), \(\mathbf{W}_{\rm opt}\), as well as \(\boldsymbol{\phi}_{\rm opt}\) and \(\boldsymbol{\alpha}_{\rm opt}\) (i.e., \(\boldsymbol{\gamma}_{\rm opt}\)). \\
\textbf{Preprocessing:}
\begin{itemize}
    \item Using the LMMSE-based RX combining matrix, reformulate \eqref{eq:R_RX_w_noise} as in \eqref{eq:Rate_linear_receiver}, and then compute the initial $\mathbf{F}$ matrix; set this matrix as \(\mathbf{F}^{(0)}\).
    \item Using the Lagrangian dual and the matrix quadratic transform,  reformulate~\eqref{eq:Rate_linear_receiver} to obtain $\overline{\mathcal{OP}}$.
\end{itemize}
\begin{algorithmic}[1]
\For{\(j = 1, 2, \ldots,J_{\rm alt}\)}
    \State Compute \(\mathbf{Y}^{(j)} =\left(\mathbf{F}^{(j-1)}\right)^{-1}\mathbf{\tilde{H}}^{(j-1)}\mathbf{V}^{(j-1)} \).
    \State Calculate \([\boldsymbol{\sigma}^{(j)}]_{i}= {\rm SINR}^{(j-1)}_i\) \(\forall i=1,\ldots,d\).
    \State Solve \eqref{eq:complement_slackness_equation} to acquire \(\lambda_{\rm opt}\) and compute:
    
    \hspace{-0.18cm}\(\mathbf{V}^{(j)}=  \bigg(\!\left(\mathbf{\tilde{H}}^{(j-1)}\right)^{\herm}\mathbf{Y}^{(j)}\mathbf{\Sigma}^{(j)}\left(\mathbf{Y}^{(j)}\right)^{\herm}\mathbf{\tilde{H}}^{(j-1)} +\) 
    
    \hspace{-0.18cm} \(\lambda_{\rm opt}\mathbf{I}_{M_T}\!\bigg)^{-1} \left(\mathbf{\tilde{H}}^{(j-1)}\right)^{\herm}\mathbf{Y}^{(j)}\mathbf{\Sigma}^{(j)}\).
    \State Compute \(\boldsymbol{\phi}^{(j)}\) via solving \(\mathcal{OP}_{\boldsymbol{\phi}}\) via RMO, and then 
    
    \hspace{-0.18cm}use this to derive the constraints \(\mathbf{l}\) and \(\mathbf{u}\).
    \State Compute \(P(\alpha_n)\cong y(\alpha_n)\) \(\forall n=1,\ldots, N\) via \eqref{eq:linear_fitting}.
    \State Set $P_{\rm RIS}^{(1)}=P_{\rm RIS}$.
    \For{\(k = 1, 2, \ldots\)}
        \State Obtain \(\boldsymbol{\alpha}^{(j)}\) via solving \(\mathcal{OP}_{\boldsymbol{\alpha}}\) with constraint $P_{\rm RIS}^{(k)}$.
        \State Set \(\boldsymbol{\gamma}^{(j)}=\boldsymbol{\phi}^{ (j)} \odot \boldsymbol{\alpha}^{(j)}\).
        \State Map the RIS reflection amplification vector \(\boldsymbol{\gamma}^{(j)}\) to 
        
         \hspace{0.348cm}its \(C_n\) and \(R_n\) parameters \(\forall n = 1,\ldots,N\) via \eqref{eq:Xn}.
        \If{\(\sum_{n=1}^N P(R_n) \leq P_{\rm RIS}\)}
            \State \textbf{Break}.
        \Else
            \State \(P_{\rm RIS}^{(k)} \gets P_{\rm RIS}^{(k)} - \sum_{n=1}^{N}\left( y(\alpha_n) - P(R_n)\right)\).
        \EndIf
    \EndFor
    \State 
    Compute \(\mathbf{F}^{(j)}\),\(\mathbf{\tilde{H}}^{(j)}\), and \(\mathcal{R}^{(j)}\) via \eqref{eq:Rate_linear_receiver}.
    \If{\(\left|\left(\mathcal{R}^{(j)} - \mathcal{R}^{(j-1)}\right)/\mathcal{R}^{(j)}\right| \leq \epsilon\)}
        \State Set \(\mathbf{V}_{\rm opt}=\mathbf{V}^{(j)}\), \(\boldsymbol{\phi}_{\rm opt}=\boldsymbol{\phi}^{(j)}\), and \(\boldsymbol{\alpha}_{\rm opt}=\boldsymbol{\alpha}^{(j)}\), 
        
        \hspace{0.348cm}and then substitute them in \eqref{eq:LMMSE_Receiver} to obtain $\mathbf{W}_{\rm opt}$.
    \EndIf
\EndFor
\State Compute \(j_{\max}=\max_{j} \mathcal{R}^{(j)}\).
\State Set \(\mathbf{V}_{\rm opt}=\mathbf{V}^{(j_{\max})}\), \(\boldsymbol{\phi}_{\rm opt}=\boldsymbol{\phi}^{(j_{\max})}\), and \(\boldsymbol{\alpha}_{\rm opt}=\boldsymbol{\alpha}^{(j_{\max})}\), and then substitute them in \eqref{eq:LMMSE_Receiver} to obtain $\mathbf{W}_{\rm opt}$.
\end{algorithmic}
\end{algorithm}

\subsection{Single-Step Decoupled Solution}\label{subsec:approximate_solution}
In this section, we propose to overlook the dependency between the \(\mathcal{OP}\)'s optimization variables to obtain a joint MIMO and active RIS Design without alternations. To this end, we focus on the high SNR regime and assume that the noise at the RX due to the active RIS is negligible with respect to the reception thermal noise. We henceforth label this approach as Decoupled Optimization (DO), and apart from being used standalone, it can serve as an initialization design (i.e., input) for the previously designed AO-based Algorithm~\ref{alg:alternate_optimization}.

\subsubsection{TX Precoder and RX Combiner Optimization}\label{subsec: precoder and combiner optimization one-shot}
Consider the Singular Value Decomposition (SVD) of the RIS-augmented MIMO channel matrix $\tilde{\mathbf{H}}=\mathbf{U}_1\boldsymbol{\Lambda}\mathbf{U}_2^{\rm H}\in\mathbb{C}^{M_R\times M_T}$, where $\mathbf{U}_2\in\mathbb{C}^{M_T\times d}$ and $\mathbf{U}_1\in\mathbb{C}^{M_R\times d}$ represent the semi-unitary matrices and $\boldsymbol{\Lambda}\in\mathbb{C}^{d\times d}$ is the diagonal matrix consisting of the $d$ non-zero singular values of $\tilde{\mathbf{H}}$ in decreasing order of magnitude. Capitalizing on the knowledge of this matrix and on our assumption herein for negligible \(\mathbf{H}_2\mathbf{\Gamma}\mathbf{n}_s\) in~\eqref{eq:system_model_antennas_streams} in comparison to $\mathbf{n}_r$, the RX combining matrix solving \(\mathcal{OP}\) is given by \(\mathbf{W}_{\rm opt} = \mathbf{U}_1\). Note that this RX linear processing cannot null the RIS-induced noise term, however, it remains nearly optimal for rate maximization when this noise is relatively weak enough~\cite{George_interference}. Furthermore, the optimal TX precoding matrix can be obtained as \(\mathbf{V}_{\rm opt} = \mathbf{U}_2 (\mathbf{P}_{\rm opt})^{\frac{1}{2}}\), where \(\mathbf{P}\triangleq{\rm diag}\{p_1,\ldots,p_d\} \in \mathbb{C}^{d\times d}\) is the per data stream power allocation, i.e., \(p_i \geq 0\) indicates the transmit power allocated to the \(i\)-th symbol in vector $\mathbf{s}$. 

The optimal power allocation matrix \(\mathbf{P}_{\rm opt}\triangleq{\rm diag}\{p_{1,{\rm opt}},\ldots,p_{d,{\rm opt}}\}\) can be obtained via the waterfilling solution~\cite{heath,George_interference}. To this end, considering the previous TX and RX designs that diagonalize \(\mathbf{\tilde{H}}\),  \eqref{eq:system_model_antennas_streams} simplifies to:
\begin{equation}\label{eq:SVD_decomp}
        \mathbf{U}_1^{\rm H}\mathbf{y}=\mathbf{\Lambda P s}+\mathbf{U}_1^{\rm H}\mathbf{H}_2\mathbf{\Gamma n}_s+\mathbf{U}_1^{\rm H}\mathbf{n}_r,
\end{equation}
indicating that $\tilde{\mathbf{H}}$ has been decomposed into \(d\) parallel eigenchannels. With $[\boldsymbol{\Lambda}]_{i,i}^2$ denoting the gain of each $i$-th eigenchannel (recall that $i=1,\ldots,d$), the instantaneous achievable spectral efficiency can be expressed as follows:
\begin{equation*}         \mathcal{R}\left(\boldsymbol{\gamma},\mathbf{P}\right)=\sum_{i=1}^{d}\log_2\!\left(1 + \underbrace{\frac{p_i [\boldsymbol{\Lambda}]_{i,i}^2}{\sigma^2F_s\mathbf{u}^{\rm H}_{1,i}\mathbf{H}_2\mathbf{\Gamma}\mathbf{\Gamma}^{\rm H}\mathbf{H}_2^{\rm H}\mathbf{u}_{1,i} +\sigma^2F_r}}_{\triangleq \overline{{\rm SINR}}_i}\right) 
\end{equation*}
with $\mathbf{u}_{1,i}$ being $\mathbf{U}_{1}$'s $i$-th column. The $\mathbf{P}_{\rm opt}$ matrix maximizing this metric subject to the constraint ${\rm Tr}\{\mathbf{P}^{\rm H}\mathbf{P}\}\leq P_{ T}$ is:
\begin{equation}
    \begin{split}
        p_{i,{\rm opt}} &=\left[\frac{1}{\eta}-\frac{1}{\rm SINR}_i\right]^{+} ,
    \end{split}
\end{equation}
where \(\eta\) is chosen to ensure that \(\sum_{i=1}^{d} p_{i,{\rm opt}} =P_T\). Note that the multitude of non-zero \(p_{i,{\rm opt}}\)'s determines the number of independent streams to be actually transmitted.


\subsubsection{RIS Phase Configuration Optimization}\label{subsec:RIS phase vector one-shot}
By performing the substitution $\mathcal{R}\left(\boldsymbol{\gamma},\mathbf{P}_{\rm out}\right)$ in the previous rate formula and considering the assumption for the negligible noise contribution from the active RIS, as compared to the thermal noise at the RX, the rate expression can be approximated as follows:
\begin{equation}\label{eq:rate_after_weak_noise_approx}
\mathcal{R}\left(\boldsymbol{\gamma}\right) \cong \sum_{i=1}^{d} \log_2\!\left(1 + \frac{p_{i,{\rm opt}} [\boldsymbol{\Lambda}]_{i,i}^2}{\sigma^2 F_r}\right).
\end{equation}

It is well known that, in the high SNR regime, equal power allocation is asymptotically optimal. This yields the following approximation and bound for the right-hand side of~\eqref{eq:rate_after_weak_noise_approx}:
\begin{align}\label{eq: approximate rate for one-shot optimization}
    &\mathcal{R}\left(\boldsymbol{\gamma}\right) \cong \sum_{i=1}^{d}\log_2\!\left(1+\frac{P_T [\boldsymbol{\Lambda}]_{i,i}^2}{d\sigma^2F_r}\right)\\
    &\le d\log_2\!\left(1+\frac{P_T\sum_{i=1}^{d}[\boldsymbol{\Lambda}]_{i,i}^2}{d^2\sigma^2F_r}\right)\!=\! d\log_2\!\left(1+\frac{P_T}{d^2\sigma^2F_r}\norm{\tilde{\mathbf{H}}}^2_{\rm F}\right),\nonumber
\end{align}
where the equality holds if and only if all eigenchannel powers $[\boldsymbol{\Lambda}]_{i,i}$'s are equal~\cite{Zhang_MIMO_IRS}. The last formula in \eqref{eq: approximate rate for one-shot optimization} indicates that the maximization of $\mathcal{R}\left(\boldsymbol{\gamma}\right)$ is equivalent to maximizing \(\norm{\tilde{\mathbf{H}}}_{\rm F}^2\). To this end, we express \(\norm{\tilde{\mathbf{H}}}_{\rm F}^2\) with respect to \(\boldsymbol{\gamma}\) as follows:
\begin{equation} \label{eq: norm of H_tilde}
\norm{\mathbf{\tilde{H}}}_{\rm F}^2\!=\!\boldsymbol{\gamma}^{\herm}\boldsymbol{\mathcal{H}}^{ \herm}\boldsymbol{\mathcal{H}}\boldsymbol{\gamma}+2\Re \left[\left(\mathrm{vec}\left\lbrace\mathbf{H}_{\rm d}\right\rbrace\right)^{\rm H}\boldsymbol{\mathcal{H}}\boldsymbol{\gamma}\right] \! + \! \norm{\mathbf{H}_{\rm d}}_{\rm F}^2.
\end{equation}
By setting \(\mathbf{T} = -\boldsymbol{\mathcal{H}}^{\herm}\boldsymbol{\mathcal{H}}\), \(\mathbf{q} = \boldsymbol{\mathcal{H}}^{\herm}\mathrm{vec}\{\mathbf{H}_{\rm d}\}\), and using the approximation~\eqref{eq:RIS_reflection coefficient vector} for \(\boldsymbol{\gamma}\), the maximization of \eqref{eq: norm of H_tilde} with respect to \(\boldsymbol{\phi}\) reduces to \(\mathcal{OP}_{\boldsymbol{\phi}}\). 
It is important to note that the optimal RIS phase configuration \(\boldsymbol{\phi}_{\rm opt}\), which can be computed using the RMO approach described previously in Section~\ref{subsec: RIS phase configuration}, is independent of the TX precoder and RX combiner matrices. In addition, we have assumed \(\boldsymbol{\bar{\alpha}} = \mathbf{1}_{N \times 1}\), which implies that all active RIS elements operate at their maximum available amplitude. This assumption is justified by the fact that \eqref{eq: norm of H_tilde} is an increasing function of \(\boldsymbol{\gamma}\)'s norm.

\subsubsection{RIS Amplitude Configuration Optimization} \label{subsec: RIS amplitude configuration (one shot)}
The maximization of~\eqref{eq: norm of H_tilde} with respect to the RIS amplitude vector \(\boldsymbol{\alpha}\) can be formulated as follows:
\begin{align*} 
\overline{\mathcal{OP}}_{\boldsymbol{\alpha}}:&\min_{\boldsymbol{\alpha}}  -\boldsymbol{\alpha}^{\tran}\mathbf{\Phi}^{\herm}\boldsymbol{\mathcal{H}}^{ \herm}\boldsymbol{\mathcal{H}}\mathbf{\Phi}\boldsymbol{\alpha} - 2{\rm Re}\left[\left(\mathrm{vec}\left\lbrace\mathbf{H}_{\rm d}\right\rbrace\right)^{\rm H}\boldsymbol{\mathcal{H}}\mathbf{\Phi}\right]\boldsymbol{\alpha}  \\
    & \,\,{\rm s.t.}\, \mathbf{l}\leq \boldsymbol{\alpha} \leq \mathbf{u},\,\,\sum_{n=1}^N y(\alpha_n)\leq P_{\rm RIS},
\end{align*}
where the boundaries \(\mathbf{l}\) and \(\mathbf{u}\) have been computed using the \(\boldsymbol{\phi}_{\rm opt}\) solution obtained in the previous step (i.e., Section~\ref{subsec:RIS phase vector one-shot}). In contrast to \(\mathcal{OP}_{\boldsymbol{\alpha}}\), $\overline{\mathcal{OP}}_{\boldsymbol{\alpha}}$ is a non-convex problem due to the fact that \(\boldsymbol{\mathcal{H}}^{ \herm}\boldsymbol{\mathcal{H}}\) is a positive semidefinite matrix. 
To overcome this and avoid alternations between \(\boldsymbol{\phi}\) and \(\boldsymbol{\alpha}\),  we resort to maximizing only the sum of the amplitudes of the active RIS unit elements, which will consequently increase \(\norm{\tilde{\mathbf{H}}}_{\rm F}^2\):
\begin{equation*}
\widetilde{\mathcal{OP}}_{\boldsymbol{\alpha}}:    \max_{\boldsymbol{\alpha}}  \sum_{n=1}^{N} \alpha_n, \,\,{\rm s.t.}\, \mathbf{l}\leq \boldsymbol{\alpha} \leq \mathbf{u},\,\,\sum_{n=1}^N y(\alpha_n)\leq P_{\rm RIS}.
\end{equation*}
Similar to our AO-based optimization framework, we follow Steps~\(6\)-\(17\) of Algorithm~\ref{alg:alternate_optimization} to design \(\boldsymbol{\alpha}\), where Step~\(9\) needs to be modified to solve $\widetilde{\mathcal{OP}}_{\boldsymbol{\alpha}}$ instead of \(\mathcal{OP}_{\boldsymbol{\alpha}}\). 
After determining \(\boldsymbol{\alpha}_{\rm opt}\), the RIS reflection amplification vector will be completely designed. Finally, the SVD of \(\mathbf{\tilde{H}}\) is computed to optimize the TX precoder and RX combiner, as described in Section~\ref{subsec: precoder and combiner optimization one-shot}, thus, completing our single-step DO approach.

\subsection{Complexity Analysis}
In this section, we investigate analytically the complexity of both presented design approaches for the considered MIMO communication system incorporating an active RIS.

\subsubsection{Single-Step DO Design}\label{subsec:Complexity one shot}
After performing \(\mathbf{\tilde{H}}\)'s SVD with the computational complexity of \(\mathcal{O}(\min\{M_T^2 M_R, M_R M_T^2\})\), \(\mathbf{V}_{\rm opt}\) is obtained in semi-closed form. The waterfilling process with respect to the resulting SINRs has complexity of \(\mathcal{O}(d^2)\), since it requires \(d\) runs for the worst case, and each run has an \(\mathcal{O}(d)\) complexity. The computation of the SINRs requires \(\mathcal{O}(M_R N)\) complexity. \(\mathbf{W}_{\rm opt}\) does not require additional computations, since it is acquired from the same SVD as \(\mathbf{V}_{\rm opt}\), thus, the cumulative complexity of acquiring \(\mathbf{V}_{\rm opt}\) and \(\mathbf{W}_{\rm opt}\) is \(\mathcal{O}\left(\min\{M_T^2 M_R, M_R M_T^2\} + d^2\right)=\mathcal{O}\left(\min\{M_T^2 M_R, M_R M_T^2\}\right)\), since \(d\leq \min\{M_T,M_R\}\). 

The RIS phase-amplitude configuration process comprises two steps. First, the Riemannian gradient descent is solved with a complexity of \(\mathcal{O}(N^{2}j_{\rm RO})\), where \(j_{\rm RO}\) denotes the number of iterations for its convergence. Secondly, $\widetilde{\mathcal{OP}}_{\boldsymbol{\alpha}}$ is a convex problem with linear constraints, hence, it can be solved with a complexity of \(\mathcal{O}\left(N^{3.5}\right)\). To account for the iterations needed to satisfy the power constraint, we introduce the value \(j_P\) which does not scale with \(N\) and, empirically, takes values in the range \([1,4]\). To conclude, the configuration of the TD-based active RIS has a total complexity of \(\mathcal{O}\left(j_{\rm RO}N^{2} + j_P N^{3.5}\right)\). Furthermore, the computation of the matrices \(\mathbf{\tilde{H}}\) and \(\boldsymbol{\mathcal{H}}\) results in the computational complexities of \(\mathcal{O}(N M_T M_R)\) and \(\mathcal{O}(N^2 M_T M_R)\), respectively. 

Putting all above together, the cumulative complexity of our single-step DO design approach proposed in~ Section~\ref{subsec:approximate_solution} is of \(\mathcal{O}\left(\min\{M_T^2 M_R, M_R M_T^2\} + j_P N^{3.5} + N^2 M_T M_R + j_{\rm RO}N^{2} \right)\), where the dominating term is \(\mathcal{O}\left(j_P N^{3.5}\right)\) due to the fact that, in practice, it is expected that \(N \gg M_T,M_R\).

\subsubsection{AO-Based Iterative Design}\label{subsec:Complexity iterative}
The \( \mathbf{V}_{\rm opt} \) with this iterative approach demands a computational complexity of \(\mathcal{O}(M_T^3 + M_T^2 M_R)\), plus the complexity of an one-dimensional bisection search, which increases logarithmically with respect to the fraction of the search interval divided by the precision. The bisection search complexity is omitted here, since all complexities are expressed without accounting for the accuracy of the arithmetic solvers. In addition, \(\mathbf{Y}_{\rm opt}\) and \(\boldsymbol{\sigma}_{\rm opt}\) require computational complexities of \(\mathcal{O}\left(M_R^3 + M_R^2 M_T\right)\) and \(\mathcal{O}\left(d M_R^3 + d M_T M_R \right)\), respectively. The RIS phase-amplitude configuration process follows steps and complexity similar to those of the DO (i.e., one shot) approach, resulting once again in \(\mathcal{O}(j_{\rm RO}N^2 + j_P N^{3.5})\). Furthermore, in each iteration, the calculation of \(\mathbf{T}\) requires \(\mathcal{O}(N^3)\) computations.

The AO-based approach iterates between the solutions for \(\mathbf{V}_{\rm opt}\), \(\mathbf{Y}_{\rm opt}\), \(\boldsymbol{\sigma}_{\rm opt}\), and \(\gamma_{\rm opt}\) for \(J_{\rm alt}\) iterations, yielding a total complexity of \(\mathcal{O}\big(J_{\rm alt} (j_P N^{3.5} + M_T^3 + dM_R^3 ) + N^2 M_T M_R \big)\), where we have kept the dominant terms. Considering that \(N \gg M_T, M_R\), the latter complexity reduces to \(\mathcal{O}(J_{\rm alt} j_P N^{3.5})\).

\section{Numerical Results and Discussion}\label{sec: Simulation Results}
In this section, we present performance evaluation results for both derived joint MIMO and active RIS designs, which are compared with carefully identified benchmarks. The efficacy of our approximate model for the RIS reflection amplification coefficient as well as the convergence properties of our AO-based iterative design are also investigated. 

\subsection{Simulation Setup}\label{subsec: Simulation Setup}
Unless otherwise indicated, we have used the following system parameters: \(M_{\rm T}=M_{\rm R}=8\) and \(d=8\), \(N=64\), $\omega = 2 \pi \times 2.4$~GHz, \(\sigma = -113.93\)~dBm (which corresponds to the thermal noise power with \(1\) MHz bandwidth at room temperature of \(20 \,{\rm C}^{\circ}\)), \(F_r = 7\)~dB, \(F_s = 5\)~dB, \(P_{\rm RIS}=1.5\) W, and \(P_T = -12.75\)~dBm, while the distance between the RIS and RX was set as \({\rm d_{RX,RIS}}=4\)~m, and that between the TX and RIS as \({\rm d_{RIS,TX}}= 40 \)~m. The parameters for the unit elements of our TD-based active RIS were chosen as follows: \(V_0=0.1\)~V, $L_1= 4.5$~nH, $L_2= 0.7$~nH, $Z_0 = 377$~$\Omega$, \(R_0 = 1.5\)~$\Omega$, \(R_n \in [-11,-1.9]\)~$\Omega$ in~\eqref{eq:R_n}, and \(C_n \in [0.85,6.25]\)~pF. For this set of parameters, yields \(\max_{\phi} \alpha_{\max}(\phi) = 30\), indicating that a single active element can potentially supply as much power as \(30\) passive elements. 
We have also considered RIS structures including mixed active and passive unit elements, with the latter having Ohmic resistance of \(1.5\)~$\Omega$. To focus on the RIS impact, we have considered the direct RX-RX MIMO channel $\mathbf{H}_{\rm d}$ completely blocked, whereas the fading conditions for the RIS-TX and RX-RIS MIMO channels $\mathbf{H}_1$ and $\mathbf{H}_2$, respectively, were drawn from the Rayleigh distribution. 

Throughout all our rate performance results, with each having been obtained after averaging over \(200\) Monte Carlo simulations, we have also used the variable $\rho\triangleq P_T {\rm P_L}/(\sigma^2 F_r)$, with \(\rm P_L \triangleq \frac{\lambda^4}{16 \pi^2} ({\rm d_{RIS,TX}}{\rm d_{RX,RIS}})^{-2}\) being the pathloss of the RX-RIS-TX cascaded channel, which actually represents the received SNR for a single-input single-output system operating under a unit-amplitude scalar channel. 
By varying the values of this parameter, the spectral efficiency behavior over different SNR values was investigated in a efficient way, instead of varying separately each of the parameters that compose it. Note that, for the previously described setting of the parameters involved in $\rho$'s definition, \(\rho = -30\)~dB is deduced. In all following results, we have set the maximum number of iterations to \(J_{\rm alt}=20\) for the proposed iterative Algorithm~\ref{alg:alternate_optimization}.

\subsection{Benchmark Schemes}\label{subsec: Bencmark Schemes}
The available literature dealing with active RIS designs often relies on simplified models neglecting phase-amplitude dependencies, uses linear power constraints, or omits the noise term contributed by the RIS~\cite{Larsson_2021,active_RIS_survey,Act_vs_Pass}. These simplifications result actually in optimization approaches that are not applicable to our case. To this end, to compare our framework with state-of-the-art optimization methods, we have utilized the Genetic Algorithm (GA) and Particle Swarm Optimization (PSO) algorithms to solve \(\mathcal{OP}\), i.e., to obtain the optimal TX precoder and the active RIS circuit parameters \(R_n, C_n\) \(\forall n=1,\ldots,N\). It is noted that the optimal LMMSE-based RX combiner was considered for both the GA and PSO algorithms. 

To evaluate \eqref{eq:Rate_linear_receiver}, both GA and PSO require a computational complexity of \(\mathcal{O}(N M_{T} M_R + d M_R^3)\), hence, their overall complexity becomes \(\mathcal{O}(K P (N M_{T} M_R + d M_R^3))\), where \(K\) represents the population size for GA or the number of particles for PSO, and \(P\) denotes the number of algorithmic iterations. To have a fair comparison with Algorithm~\ref{alg:alternate_optimization} in terms of asymptotic complexity, we have set \(K\) and \(P\) in our comparisons so that \(\mathcal{O}(K P (N M_{T} M_R + d M_R^3))\) matches the cumulative complexity of our AO-based design, as presented in Section~\ref{subsec:Complexity iterative}. 

We have also introduced the Phase-Amplitude Independent Decoupled Optimization (PAIDO) benchmark, which applies our single-step DO framework without leveraging the phase-amplitude dependence formula in~\eqref{eq:RIS_reflection coefficient vector}. Instead, this simplification treats \(\boldsymbol{\alpha}\) and \(\boldsymbol{\phi}\) as independent variables maximizing \eqref{eq: norm of H_tilde} with respect to \(\boldsymbol{\phi}\) by substituting \(\boldsymbol{\gamma} = \boldsymbol{\alpha} \odot \boldsymbol{\phi}\). The comparisons between DO and PAIDO highlight the amount of performance degradation when the phase-amplitude dependence is neglected. 

\subsection{Convergence of the AO-Based Iterative Design}
\begin{figure}[t]
    \centering
    \includegraphics[width=\columnwidth]{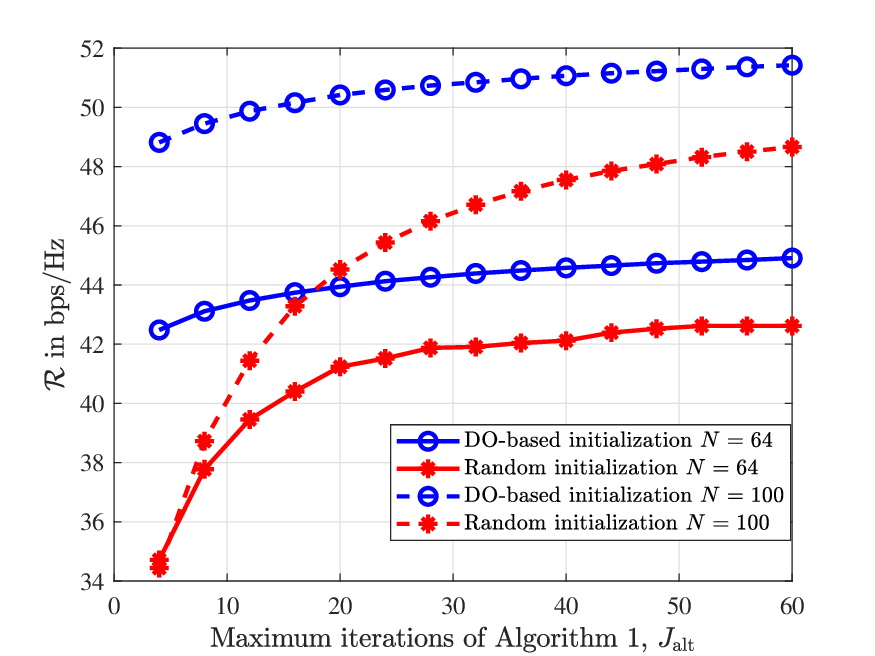}
    \caption{The convergence behavior of Algorithm~\ref{alg:alternate_optimization} under two different initialization strategies: 1) via the DO framework, and 2) random initialization; the achievable spectral efficiency is illustrated versus increasing \(J_{\rm alt}\) values. The dashed and solid lines correspond to \(N=64\) with \(P_{\rm RIS}=1.5\) W and \(N=100\) with \(P_{\rm RIS}=2.32\) W, respectively. }
    \label{fig: convergence}
\end{figure}

We commence with the investigation of the convergence behavior of the proposed Algorithm~\ref{alg:alternate_optimization} in Fig.~\ref{fig: convergence}. In particular, the achievable spectral efficiency utilizing Algorithm~\ref{alg:alternate_optimization} for joint MIMO and active RIS design is plotted versus the number of maximum iterations \(J_{\rm alt}\), considering two different initializations (i.e., \(\mathbf{V}^{(0)}\), \(\boldsymbol{\phi}^{(0)}\), and \(\boldsymbol{\alpha}^{(0)}\)): one based on the proposed single-step DO design (blue curve with circles) and the other on a random setting (red curve with asterisks). To showcase the convergence behavior with respect to the number of RIS elements as well, we investigate two cases for each initialization regime: \(N=64\) active elements with \(P_{\rm RIS}=1.5\) W and \(N=100\) with \(P_{\rm RIS}=2.32\) W. The power constraints were set so that \(P_{\rm RIS}=0.8NP(-11)\), where \(P(-11)\) is the maximum power consumption of a single unit element. Consequently, not all RIS elements can operate at their peak power levels, necessitating a power distribution at the entire RIS panel. It can be seen in the figure that, not only the convergence behavior is enhanced with the proposed DO-based initialization, but also the rate performance. In fact, even in the extreme case where Algorithm~\ref{alg:alternate_optimization} runs with DO-based initialization for \(J_{\rm alt}=4\) iterations, the achievable spectral efficiency is comparable to that with random initialization for \(J_{\rm alt}=60\) iterations, for both cases of \(N=64\) and \(100\). Furthermore, for up to \(J_{\rm alt}=16\) iterations, the DO-based initialization with \(N=64\) elements achieves even larger spectral efficiency values than the randomly initialized case with \(N=100\), despite having less active elements and a lower power budget.

\subsection{Achievable Spectral Efficiency Results}
\begin{figure}[t]
    \centering
    \includegraphics[width=\columnwidth]{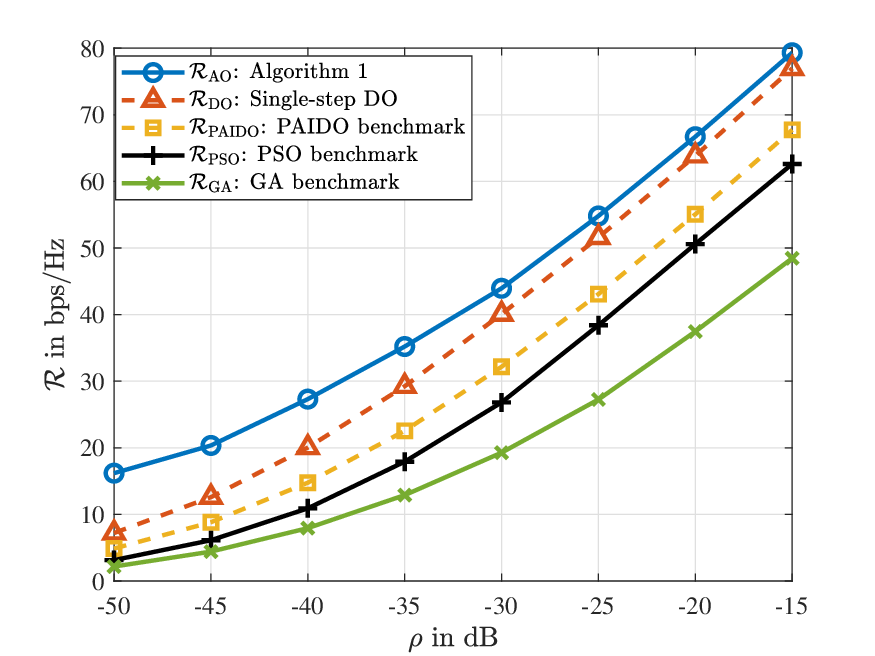}
    \caption{Achievable spectral efficiency versus the different \(\rho\) values resulting from the TX power levels \(P_T \in [-32.75,22.75]\)~dBm, considering both proposed designs and all benchmark schemes described in Section~\ref{subsec: Bencmark Schemes}. 
    }
    \label{fig: rate vs SNR}
\end{figure}

The achievable spectral efficiency with respect to different \(\rho\) values for both the proposed joint MIMO and active RIS designs (\(\mathcal{R}_{\rm AO}\) for Algorithm~\ref{alg:alternate_optimization} with DO-based initialization and \(\mathcal{R}_{\rm DO}\)) and all benchmarks described in Section~\ref{subsec: Simulation Setup} (\(\mathcal{R}_{\rm PAIDO}\), \(\mathcal{R}_{\rm PSO}\), and \(\mathcal{R}_{\rm GA}\)) is illustrated in Fig.~\ref{fig: rate vs SNR}. It can be observed that the proposed AO-based and single-step DO designs consistently outperform PSO and GA across all \(\rho\) values. Notably, the DO design requires asymptotic computational complexity of around \(\mathcal{O}(j_{\rm P} N^{3.5})\), being lower than that of Algorithm~\ref{alg:alternate_optimization}, which is \(\mathcal{O}(J_{\rm alt} j_{\rm P} N^{3.5})\), as well as that of the PSO and GA schemes which was fixed to the latter value. As also shown, \(\mathcal{R}_{\rm DO}\) converges to \(\mathcal{R}_{\rm AO}\) as \(\rho\) increases. This is attributed to the fact that the approximations used in the DO to decouple $\mathcal{OP}$ are sufficiently tight in the high SNR regime and when the RIS-induced noise at the RX is negligible with respect to
the thermal noise. In addition, it is evident that DO significantly outperforms PAIDO, which ignores the phase-amplitude dependence at the active RIS unit elements. In fact, \(\mathcal{R}_{\rm PAIDO}\) does not converge to \(\mathcal{R}_{\rm AO}\) at high SNR values, as actually \(\mathcal{R}_{\rm DO}\) does, indicating that neglecting the dependence between the phase and the amplitude creates a constant performance gap. In more details, PAIDO fails to capture the dependency of the amplitude intervals on the phase values, thus, limiting the feasible amplitude range and, consequently, degrading performance. Nevertheless, PAIDO outperforms both PSO and GA for all simulated values of \(\rho\), closely following PSO performance across the entire \(\rho\) range.

\begin{figure}[t]
    \centering
    \includegraphics[width=\columnwidth]{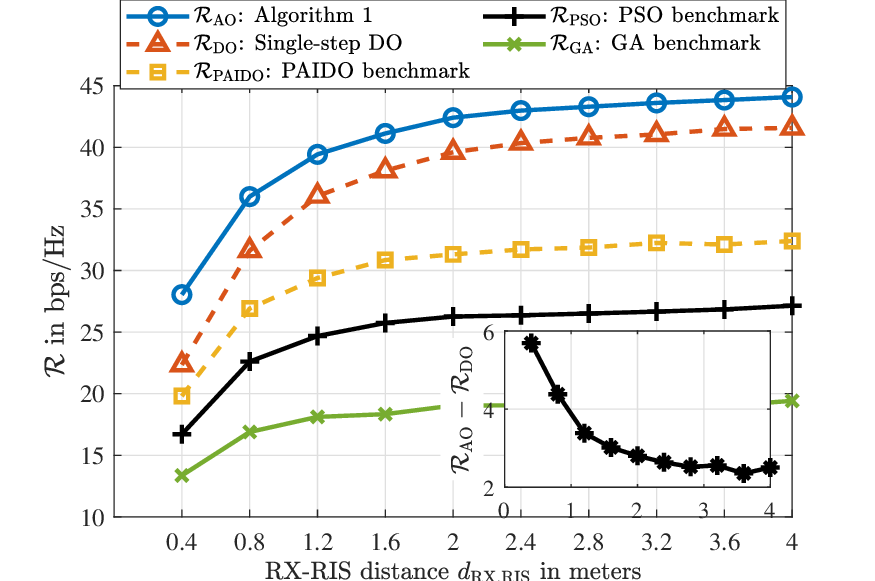}
    \caption{Achievable spectral efficiency with respect to different \({\rm d_{RX,RIS}}\) values for \(\rho=-30\)~dB, considering both proposed designs and all benchmark schemes described in Section~\ref{subsec: Simulation Setup}. Decreasing \({\rm d_{RX,RIS}}\) increases the amplitude of \(\mathbf{H}_2\), thus, intensifying the noise contribution from the active RIS. 
    The inset figure (with the same \({\rm d_{RX,RIS}}\) values in the horizontal axis) depicts the difference \(\mathcal{R}_{\rm AO} - \mathcal{R}_{\rm DO}\), showcasing the role of the RIS-induced noise neglection; the proposed AO-based iterative design takes this noise under consideration, whereas the proposed single-step DO design ignores it.}
    \label{fig: rate vs distance}
\end{figure}

The considered TD-based active RIS contributes the noise terms \(\sigma^2F_s\norm{\mathbf{w}_i \mathbf{H}_2 \boldsymbol{\Gamma}}_{\rm F}^2\) \(\forall i=1,\ldots,d\) appearing in the spectral efficiency formula in~\eqref{eq:R_RX_w_noise}. As previously discussed, our iterative AO-based design accounts for this noise when designing \(\mathbf{V}\), \(\mathbf{W}\), and \(\boldsymbol{\gamma}\), while our DO design considers it only during \(\mathbf{V}\)'s design and ignores it for obtaining \(\mathbf{W}\) and \(\boldsymbol{\gamma}\). In Fig.~\ref{fig: rate vs distance}, the achievable spectral efficiency with all considered schemes is plotted as a function of the \({\rm d_{RX,RIS}}\) distance. Note that, when \({\rm d_{RX,RIS}}\) decreases, \(\mathbf{H}_2\)'s gain increases thereby amplifying the RIS-induced noise. To focus on the investigation of the impact of this noise, the SNR was kept constant to \(\rho=-30\)~dB, compensating for changes in \(\mathbf{H}_2\)'s pathloss with inverse changes in \(\mathbf{H}_1\) to maintain constant cascaded pathloss. The results demonstrate that our AO-based design consistently outperforms all benchmarks across all distances. It can be seen that, for \({\rm d_{RX,RIS}} \geq 2.4\) m where the RIS-induced noise becomes negligible, all schemes' rates saturate indicating that further RX-RIS distances have minimal effect on performance.

The inset figure in Fig.~\ref{fig: rate vs distance} depicts the difference \(\mathcal{R}_{\rm AO} - \mathcal{R}_{\rm DO}\), i.e., the difference between the achievable spectral efficiencies with Algorithm~\ref{alg:alternate_optimization} and the DO design. As expected, this difference decreases with increasing \({\rm d_{RX,RIS}}\), eventually saturating around \(2\) bps/Hz, which is consistent with Fig.~\ref{fig: rate vs SNR}. It is also shown that the PAIDO benchmark fails to converge to our AO-based design, while significantly underperforming our single-step DO one. This behavior reaffirms the importance of considering the phase-amplitude dependency in the active RIS optimization, a crucial aspect often overlooked in the literature. It can be finally observed that \(\mathcal{R}_{\rm PSO}\) and \(\mathcal{R}_{\rm GA}\) exhibit inferior performance compared to AO, DO, and PAIDO, underscoring the superiority of our framework as well as its potential for near-optimal performance in multi-stream MIMO systems incorporating a TD-based active RIS.

\begin{figure}[t]
    \centering
    \includegraphics[width=\columnwidth]{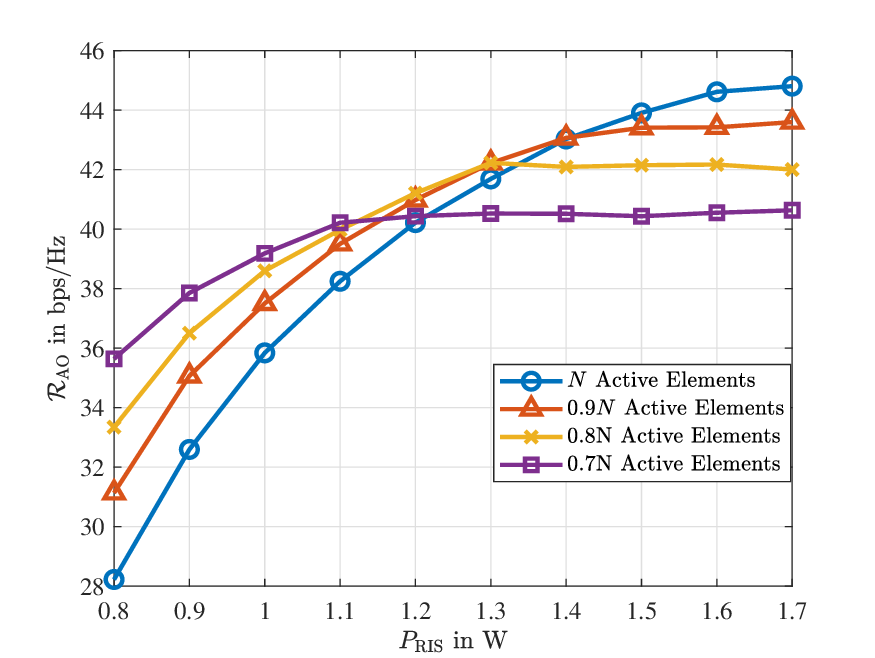}
    \caption{Achievable spectral efficiency \(\mathcal{R}_{\rm AO}\) versus \(P_{\rm RIS}\) for varying numbers of active unit elements, considering an RIS with $N=64$ mixed active and passive elements. Note that, to meet the overall RIS power budget \(P_{\rm RIS}\), the more are the active unit elements, the less is their available maximum amplitude and its range.}
    \label{fig: rate vs RIS power}
\end{figure}

For the system parameters described in Section~\ref{subsec: Simulation Setup}, the power consumption per RIS unit element ranges from \(P(-1.9) = 8\) mW (minimum) to \(P(-11) = 26.5\) mW (maximum). Thus, the total power consumption alters between \(0.5\) W and \(1.7\) W for the considered RIS with \(N=64\) active elements, which is a reasonable value for practical applications. Figure~\ref{fig: rate vs RIS power} depicts the achievable spectral efficiency using Algorithm~\ref{alg:alternate_optimization} as a function of the active RIS power budget \(P_{\rm RIS} \in [0.8,1.7]\)~W, capturing scenarios from tight to relaxed power constraints. We have considered an RIS with all its $N$ identical unit elements being active, as well RIS structures with mixed active and passive elements via random assignments. 
For the passive unit elements, the amplitude configuration stage of our AO-based design has been omitted, since control over this parameter is infeasible, and the corresponding element modeling was adjusted as described in the remark of Section~\ref{remark: active and passive unit cells}. It can be observed from the figure that maintaining all \(N\) elements active yields suboptimal performance, especially under tight power constraints. This behavior indicates that the power is inefficiently distributed across the unit elements. Specifically, the configuration with \(0.7N\) active unit elements (i.e., the one with the most passive elements) outperforms all others for \(P_{\rm RIS} \leq 1.1\)~W. For \(1.1 \,\text{W} \leq P_{\rm RIS} \leq 1.3\,\text{W}\), the \(0.8N\) case is superior, while \(0.9N\) performs better for \(1.3 \,\text{W} \leq P_{\rm RIS} \leq 1.4\,\text{W}\). All elements being active gives the best rate performance when \(P_{\rm RIS} \geq 1.4\)~W. 

The results in Fig.~\ref{fig: rate vs RIS power} indicate that, even with a random assignment on the RIS unit elements to be passive, significant performance improvements are obtained, with this improvement being dependent on the overall RIS power budget. By switching some elements to passive reflection (i.e., only phase configuration), power can be reallocated to active elements, enabling them to achieve higher amplitudes. For instance, unit elements with limited amplitude range due to their applied phase shifts could switch to passive mode, freeing up power for elements that can reach larger amplitudes. Optimizing this selection would involve solving a discrete optimization problem to determine which elements should operate in active or passive modes based on RIS power constraints and channel conditions, a promising direction for future research.

\begin{figure}[t]
    \centering
    \includegraphics[width=\columnwidth]{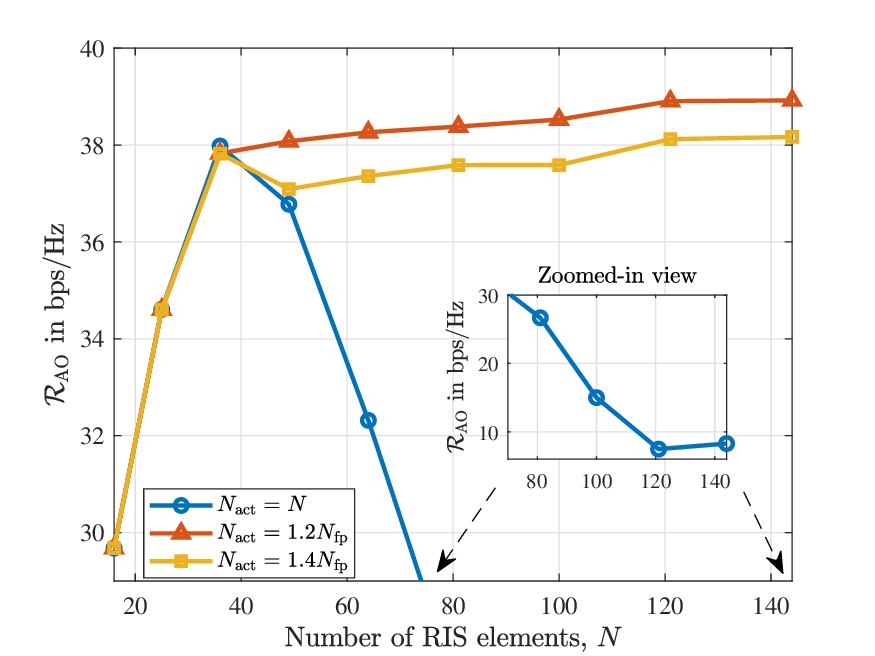}
    \caption{Achievable spectral efficiency \(\mathcal{R}_{\rm AO}\) versus the number of RIS elements \(N\) considering the fixed power budget \(P_{\rm RIS}=0.9\) W. For this power constraint, \(N_{\rm fp}=34\) elements can oparate with the maximum power consumption, since \(P(-11)=26.5\) mW is deduced. Three performance curves are shown, each for a different number of active elements: \textit{i}) all elements are active \(N_{\rm act}=N\); \textit{ii}) the number of active elements are \(N_{\rm act}=1.2N_{\rm fp}=40\) for \(N\geq 49\) and \(N_{\rm act}=N\) for \(N < 49\); and \textit{iii}) the number of active elements are \(N_{\rm act}=1.4 N_{\rm fp}=47\) for \(N\geq 49\) and \(N_{\rm act}=N\) for \(N < 49\). The inset figure shows the curve for \(N_{\rm act}=N\) when \(N\) takes values within \([81,144]\).
    }
    \label{fig: rate vs RIS elements}
\end{figure}

Figure~\ref{fig: rate vs RIS elements} depicts the \(\mathcal{R}_{\rm AO}\) rate performance as a function of the total number of RIS elements $N$ under a fixed power budget of \(P_{\rm RIS} = 0.9\)~W. With this constraint, up to \(N_{\rm fp} = 34\) elements can operate at full power (since \(34P(-11) = 0.9\) W), while, for \(N > N_{\rm fp}\), the power is redistributed. We consider three configurations: \textit{i}) all elements are active (\(N_{\rm act} = N\)); \textit{ii}) \(N_{\rm act} = 1.2N_{\rm fp} = 40\) for \(N \geq 49\) and \(N_{\rm act} = N\) otherwise; and \textit{iii}) \(N_{\rm act} = 1.4N_{\rm fp} = 47\) for \(N \geq 49\), otherwise, \(N_{\rm act} = N\). It can be observed that, for \(N_{\rm act} = N\) (circle markers), \(\mathcal{R}_{\rm AO}\) increases up to the \(N = 36\) case, but degrades significantly for larger \(N\) due to inefficient power distribution. The performance gets stabilized around \(N=121\), after which the rate improves slightly, as shown in the zoomed-in region in the inset figure. This is attributed to the fact that the maximum number of elements that can operate actively with \(P_{\rm RIS}=0.9\)~W is \(N=120\), since \(120P(-1.9) = 0.9\) W. From that point on, any additional elements are switched to passive mode. Consequently, the performance saturates at \(N = 144\), since power cannot be overdistributed more, and passive elements are added which provide marginal improvements compared to the \(N=121\) case. 

To further illustrate why overdistributing the available power budget is inefficient, we now investigate the achieved reflection amplitude. At minimal power consumption, i.e., \(P(-1.9) = 8\)~mW, the maximum reflection amplitude is \(\max_{\varphi} \alpha_{\min}(\varphi) = 1.38\), whereas, at full power, i.e., \(P(-11) = 26.5\)mW, it reaches \(\max_{\varphi} \alpha_{\max}(\varphi) = 30\). In contrast, passive elements (with \(1.5\,\Omega\) resistance) achieve \(\max_{\varphi} \alpha_{\rm pass}(\varphi) = 0.99\), indicating that elements of minimal power provide only slight gains over passive ones. Notably, three minimal-power active RIS elements consume nearly the same power as a single full-power element, yet the latter is significantly more efficient, offering approximately \(22\) times larger maximum amplitude. This efficiency disparity is evident in the rate performance, where even \(16\) fully powered active elements outperform \(121\) minimal-powered ones.

To determine the optimal number of active elements under a fixed power budget, we have considered thresholds based on \(N_{\rm fp}\), testing cases where \(N_{\rm act} = 1.2N_{\rm fp}\) (red triangles) and \(N_{\rm act} = 1.4N_{\rm fp}\) (yellow squares). As shown, the \(1.2N_{\rm fp}\) configuration achieves the best performance, with that of \(1.4N_{\rm fp}\) following closely, but being consistently lower. This suggests that an effective strategy is to determine \(N_{\rm act}\) based on \(N_{\rm fp}\), with \(1.2N_{\rm fp}\) offering a promising balance between power allocation and achievable rate performance. These findings highlight the importance of optimizing the trade-off between active and passive RIS elements under power constraints, which remains a promising avenue for future work.

\section{Conclusions and Future Work}\label{sec: Conclusion}
This paper studied circuit-level modeling of active RIS unit elements realized with TDs, deriving novel realistic amplification limits and a power consumption model directly from the diode's I-V characteristics. 
Unlike prior works, which often impose arbitrary amplitude constraints or use linear amplifier-based models, we incorporated practical phase-amplitude dependencies and circuit constraints to ensure feasibility. We studied a multi-stream MIMO communication system assisted by a TD-based active RIS and introduced an AO-based framework to jointly design the TX precoder, RX combiner, and RIS reflection amplification configuration to maximize the achievable spectral efficiency while adhering to circuit-related constraints. Additionally, we proposed a single-step initialization method that improves both convergence and performance, achieving near-optimal results under high SNRs and weak RIS-induced noise conditions.
Our numerical investigations showcased the importance of accurately modeling the phase-amplitude dependence of active RIS unit elements, as well as the RX-RIS distances above which the RIS-induced noise can be neglected thanks to pathloss. It was also demonstrated that, increasing the number of active elements on a fixed sized RIS structure with mixed active and passive elements, does not improve the achievable rate at low RIS power budget levels.  
A promising future direction involves the optimization of the assignment of RIS elements to active or passive, which will also depend on system specifications and channel conditions. An even more holistic approach would be to dynamically optimize the assignment between: 1) joint amplitude and phase control, 2) phase control only (with power consumption), and 3) static metamaterials (with no power consumption).

This work corroborated the potential of active RISs with reflection amplification capabilities to overcome key limitations of conventional RISs with phase-only reconfigurability, making them a viable solution~\cite{GR_001} for non line-of-sight channel conditions, extended cell-edge coverage, and high-frequency bands (millimeter waves and THz) where severe pathloss occurs. They also enable interference suppression by selectively amplifying desired signals while attenuating interference, enhance security by improving legitimate users' SNR while reducing the same for eavesdroppers, and promise effectiveness in challenging future high-frequency communications scenarios~\cite{ETSI_THz_2023}, such as wireless X-haul transport, data centers interconnection, as well as in-vehicle and indoor wireless connectivity~\cite{AJG2024}.

\bibliographystyle{IEEEtran}
\bibliography{references}

\begin{thebibliography}{10}
\providecommand{\url}[1]{#1}
\csname url@samestyle\endcsname
\providecommand{\newblock}{\relax}
\providecommand{\bibinfo}[2]{#2}
\providecommand{\BIBentrySTDinterwordspacing}{\spaceskip=0pt\relax}
\providecommand{\BIBentryALTinterwordstretchfactor}{4}
\providecommand{\BIBentryALTinterwordspacing}{\spaceskip=\fontdimen2\font plus
\BIBentryALTinterwordstretchfactor\fontdimen3\font minus \fontdimen4\font\relax}
\providecommand{\BIBforeignlanguage}[2]{{%
\expandafter\ifx\csname l@#1\endcsname\relax
\typeout{** WARNING: IEEEtran.bst: No hyphenation pattern has been}%
\typeout{** loaded for the language `#1'. Using the pattern for}%
\typeout{** the default language instead.}%
\else
\language=\csname l@#1\endcsname
\fi
#2}}
\providecommand{\BIBdecl}{\relax}
\BIBdecl

\bibitem{George_Basar_RIS_magazine}
E.~Basar \emph{et~al.}, ``Reconfigurable intelligent surfaces for {6G}: Emerging hardware architectures, applications, and open challenges,'' \emph{IEEE Veh. Technol. Mag.}, vol.~19, no.~3, pp. 27--47, 2024.

\bibitem{EURASIP_RIS}
G.~C. Alexandropoulos \emph{et~al.}, ``{RIS}-enabled smart wireless environments: {D}eployment scenarios, network architecture, bandwidth and area of influence,'' \emph{EURASIP J. Wireless Commun. Netw.}, vol. 103, pp. 1--38, Oct. 2023.

\bibitem{9864655}
------, ``Pervasive machine learning for smart radio environments enabled by reconfigurable intelligent surfaces,'' \emph{Proc. IEEE}, vol. 110, no.~9, pp. 1494--1525, 2022.

\bibitem{WavePropTCCN}
------, ``Reconfigurable intelligent surfaces and metamaterials: {T}he potential of wave propagation control for {6G} wireless communications,'' \emph{IEEE ComSoc TCCN Newslett.}, vol.~6, no.~1, pp. 25--37, 2020.

\bibitem{Marco_Visionary_2019}
M.~Di~Renzo \emph{et~al.}, ``{Smart radio environments empowered by {AI} reconfigurable meta-surfaces: An idea whose time has come},'' \emph{EURASIP J. Wireless Commun. Netw.}, vol. 129, pp. 1--20, 2019.

\bibitem{HMIMO}
C.~Huang \emph{et~al.}, ``Holographic {MIMO} surfaces for {6G} wireless networks: {O}pportunities, challenges, and trends,'' \emph{IEEE Wireless Commun.}, vol.~27, no.~5, pp. 118--125, 2020.

\bibitem{10247149}
R.~Singh \emph{et~al.}, ``Indexed multiple access with reconfigurable intelligent surfaces: The reflection tuning potential,'' \emph{IEEE Commun. Mag.}, vol.~62, no.~4, pp. 120--126, 2024.

\bibitem{9217944}
S.~Lin \emph{et~al.}, ``Reconfigurable intelligent surfaces with reflection pattern modulation: Beamforming design and performance analysis,'' \emph{IEEE Trans. Wireless Commun.}, vol.~20, no.~2, pp. 741--754, 2021.

\bibitem{ibrahim2022binary}
E.~Ibrahim \emph{et~al.}, ``Binary polarization shift keying with reconfigurable intelligent surfaces,'' \emph{IEEE Wireless Commun. Lett.}, vol.~11, no.~5, pp. 908--912, 2022.

\bibitem{8064681}
K.~T. Pham \emph{et~al.}, ``Dual-band transmitarrays with dual-linear polarization at {Ka}-band,'' \emph{IEEE Trans. Ant. Propag.}, vol.~65, no.~12, pp. 7009--7018, 2017.

\bibitem{10693440}
A.~Ghaneizadeh \emph{et~al.}, ``Metasurface energy harvesters: State-of-the-art designs and their potential for energy sustainable reconfigurable intelligent surfaces,'' \emph{IEEE Access}, vol.~12, pp. 160\,464--160\,494, 2024.

\bibitem{10243495}
S.~P. Chepuri \emph{et~al.}, ``Integrated sensing and communications with reconfigurable intelligent surfaces: From signal modeling to processing,'' \emph{IEEE Signal Process. Mag.}, vol.~40, no.~6, pp. 41--62, 2023.

\bibitem{Jian_RIS_survey}
M.~Jian \emph{et~al.}, ``Reconfigurable intelligent surfaces for wireless communications: Overview of hardware designs, channel models, and estimation techniques,'' \emph{Intell. Converged Netw.}, vol.~3, no.~1, pp. 1--32, 2022.

\bibitem{RIS_prototypes_and_power_consumption_model}
J.~Wang \emph{et~al.}, ``Reconfigurable intelligent surface: Power consumption modeling and practical measurement validation,'' \emph{IEEE Trans. Commun.}, vol.~72, no.~9, pp. 5720--5734, 2024.

\bibitem{abadal2020programmable}
S.~Abadal \emph{et~al.}, ``Programmable metamaterials for software-defined electromagnetic control: Circuits, systems, and architectures,'' \emph{IEEE J. Emerg. Sel. Topics Circuits Syst.}, vol.~10, no.~1, pp. 6--19, 2020.

\bibitem{amri2022recent}
M.~M. Amri, ``Recent trends in the {R}econfigurable {I}ntelligent {S}urfaces ({RIS}): Active {RIS} to brain-controlled {RIS},'' in \emph{Proc. IEEE COMNETSAT}, Solo, Indonesia, 2022.

\bibitem{Act_vs_Pass}
Z.~Zhang \emph{et~al.}, ``Active {RIS} vs. passive {RIS}: Which will prevail in 6{G}?'' \emph{IEEE Trans. Commun.}, vol.~71, no.~3, pp. 1707--1725, 2023.

\bibitem{MIMO_active_RIS_robust_BF}
R.~Allu \emph{et~al.}, ``Robust beamformer design in active ris-assisted multiuser mimo cognitive radio networks,'' \emph{IEEE Trans. Cogn. Commun. Netw.}, vol.~9, no.~2, pp. 398--413, 2023.

\bibitem{active_RIS_survey}
M.~Ahmed \emph{et~al.}, ``Active reconfigurable intelligent surfaces: Expanding the frontiers of wireless communication-{A} survey,'' \emph{IEEE Commun. Surveys Tuts.}, early, 2024.

\bibitem{Active_Scatterer_2012}
J.~Bousquet \emph{et~al.}, ``A {4-GHz} active scatterer in \(130\)-nm {CMOS} for phase sweep amplify-and-forward,'' \emph{IEEE Trans. Circuits Syst. I}, vol.~59, no.~3, pp. 529--540, 2012.

\bibitem{Tunneling_RFID_2018}
F.~Amato \emph{et~al.}, ``Tunneling {RFID} tags for long-range and low-power microwave applications,'' \emph{IEEE J. Radio Freq. Identif.}, vol.~2, no.~2, pp. 93--103, 2018.

\bibitem{Tag_Backscatter_2014}
J.~Kimionis \emph{et~al.}, ``Enhancement of {RF} tag backscatter efficiency with low-power reflection amplifiers,'' \emph{IEEE Trans. Microw. Theory Tech.}, vol.~62, no.~12, pp. 3562--3571, 2014.

\bibitem{Active_Backscatter_2019}
S.~Khaledian \emph{et~al.}, ``Active two-way backscatter modulation: {A}n analytical study,'' \emph{IEEE Trans. Wireless Commun.}, vol.~18, no.~3, pp. 1874--1886, 2019.

\bibitem{Larsson_2021}
R.~Long \emph{et~al.}, ``Active reconfigurable intelligent surface-aided wireless communications,'' \emph{IEEE Trans. Wireless Commun.}, vol.~20, no.~8, pp. 4962--4975, 2021.

\bibitem{Power_Budget}
K.~Zhi \emph{et~al.}, ``Active {RIS} versus passive {RIS}: Which is superior with the same power budget?'' \emph{IEEE Commun. Lett.}, vol.~26, no.~5, pp. 1150--1154, 2022.

\bibitem{8369144}
D.~Mishra \emph{et~al.}, ``Energy sustainable {IoT} with individual {QoS} constraints through {MISO} {SWIPT} multicasting,'' \emph{IEEE Internet of Things J.}, vol.~5, no.~4, pp. 2856--2867, 2018.

\bibitem{9501003}
G.~C. Alexandropoulos \emph{et~al.}, ``Safeguarding {MIMO} communications with reconfigurable metasurfaces and artificial noise,'' in \emph{Proc. IEEE ICC}, Montreal, Canada, 2021.

\bibitem{SWIPT}
Y.~Gao \emph{et~al.}, ``Beamforming optimization for active intelligent reflecting surface-aided {SWIPT},'' \emph{IEEE Trans. Wireless Commun.}, vol.~22, no.~1, pp. 362--378, 2023.

\bibitem{zeng2022throughput}
P.~Zeng \emph{et~al.}, ``Throughput maximization for active intelligent reflecting surface-aided wireless powered communications,'' \emph{IEEE Wireless Commun. Lett.}, vol.~11, no.~5, pp. 992--996, 2022.

\bibitem{SecrecyBai}
L.~Dong, H.-M. Wang, and J.~Bai, ``Active reconfigurable intelligent surface aided secure transmission,'' \emph{IEEE Trans. Veh. Technol.}, vol.~71, no.~2, pp. 2181--2186, 2022.

\bibitem{Khoshafa_2021}
M.~H. Khoshafa \emph{et~al.}, ``Active reconfigurable intelligent surfaces-aided wireless communication system,'' \emph{IEEE Commun. Lett.}, vol.~25, no.~11, pp. 3699--3703, Nov. 2021.

\bibitem{Secrecy_Green}
W.~Lv \emph{et~al.}, ``{RIS}-assisted green secure communications: Active {RIS} or passive {RIS}?'' \emph{IEEE Wireless Commun. Lett.}, vol.~12, no.~2, pp. 237--241, 2023.

\bibitem{10636047}
Q.~Li \emph{et~al.}, ``Cooperative backscatter communications with reconfigurable intelligent surfaces: An {APSK} approach,'' \emph{IEEE Trans. Wireless Commun.}, vol.~23, no.~11, pp. 16\,218--16\,233, 2024.

\bibitem{optimization_of_number_of_active_elements}
S.~Ahmed \emph{et~al.}, ``Adding active elements to reconfigurable intelligent surfaces to enhance energy harvesting for {IoT} devices,'' in \emph{Proc. IEEE MILCOM}, San Diego, USA, 2021.

\bibitem{HybrAct}
N.~T. Nguyen \emph{et~al.}, ``Hybrid active-passive reconfigurable intelligent surface-assisted multi-user {MISO} systems,'' in \emph{Proc. IEEE SPAWC}, Oulu, Finland, 2022.

\bibitem{SubConActive}
K.~Liu \emph{et~al.}, ``Active reconfigurable intelligent surface: Fully-connected or sub-connected?'' \emph{IEEE Commun. Lett.}, vol.~26, no.~1, pp. 167--171, 2022.

\bibitem{altunbasAct}
Z.~Yigit \emph{et~al.}, ``Hybrid reflection modulation,'' \emph{IEEE Trans. Wireless Commun.}, pp. 1--1, 2022.

\bibitem{George_Active}
R.~A. Tasci \emph{et~al.}, ``A new {RIS} architecture with a single power amplifier: Energy efficiency and error performance analysis,'' \emph{IEEE Access}, vol.~10, pp. 44\,804--44\,815, 2022.

\bibitem{Abeywickrama_2020}
S.~Abeywickrama \emph{et~al.}, ``Intelligent reflecting surface: {P}ractical phase shift model and beamforming optimization,'' \emph{IEEE Trans. Commun.}, vol.~68, no.~9, pp. 5849--5863, 2020.

\bibitem{cui2014coding}
T.~J. Cui \emph{et~al.}, ``Coding metamaterials, digital metamaterials and programmable metamaterials,'' \emph{Light: Sci. \& Appl.}, vol.~3, no.~10, p. e218, 2014.

\bibitem{PhysRevApplied.11.044024}
F.~Liu \emph{et~al.}, ``Intelligent metasurfaces with continuously tunable local surface impedance for multiple reconfigurable functions,'' \emph{Phys. Rev. Applied}, vol.~11, p. 044024, 2019.

\bibitem{amp1}
R.~Taniloo, ``Negative power devices and absolute negative resistance: Their energy behavior,'' in \emph{Proc. IEEE Norchip Conf.}, Linköping, Sweden, 2006.

\bibitem{8057976}
S.~{Khaledian} \emph{et~al.}, ``A full-duplex bidirectional amplifier with low {DC} power consumption using tunnel diodes,'' \emph{IEEE Microw. Wireless Compon. Lett}, vol.~27, no.~12, pp. 1125--1127, 2017.

\bibitem{nonlinear}
J.~Lee and K.~Yang, ``{RF} power analysis on \(5.8\) {GHz} low-power amplifier using resonant tunneling diodes,'' \emph{IEEE IEEE Microw. Wireless Compon. Lett.}, vol.~27, no.~1, pp. 61--63, 2016.

\bibitem{sze_semiconductor}
S.~M. Sze, \emph{Semiconductor devices: physics and technology}.\hskip 1em plus 0.5em minus 0.4em\relax John wiley \& sons, 2008.

\bibitem{BERGER2011176}
P.~Berger and A.~Ramesh, \emph{5.05 - Negative Differential Resistance Devices and Circuits}, P.~Bhattacharya, R.~Fornari, and H.~Kamimura, Eds.\hskip 1em plus 0.5em minus 0.4em\relax Elsevier, 2011.

\bibitem{modelling_patent}
\BIBentryALTinterwordspacing
J.~Bi \emph{et~al.}, ``Multi - layered tunnel junction structure, light emitting device having the same, and production method of such device,'' Patent US 11,011,674 B2, May 18, 2021. [Online]. Available: \url{https://patents.google.com/patent/US11011674B2/en}
\BIBentrySTDinterwordspacing

\bibitem{kriplani2011modelling}
N.~M. Kriplani \emph{et~al.}, ``Modelling of an {E}saki tunnel diode in a circuit simulator,'' \emph{Hindawi Act. Passive Electron. Compon.}, vol. 2011, 2011.

\bibitem{Lambert_function}
R.~Iacono and J.~P. Boyd, ``New approximations to the principal real-valued branch of the {L}ambert {W}-function,'' \emph{Adv. Comput. Mathematics}, vol.~43, no.~6, pp. 1403--1436, 2017.

\bibitem{George_RIS_TWC2019}
C.~Huang \emph{et~al.}, ``{Reconfigurable intelligent surfaces for energy efficiency in wireless communication},'' \emph{IEEE Trans. Wireless Commun.}, vol.~18, no.~8, pp. 4157--4170, 2019.

\bibitem{6047578}
J.-F. Bousquet \emph{et~al.}, ``A \(4\)-{GHz} active scatterer in \(130\)-nm {CMOS} for phase sweep amplify-and-forward,'' \emph{IEEE Trans. Circuits Sys. I}, vol.~59, no.~3, pp. 529--540, 2012.

\bibitem{heath}
R.~W. Heath~Jr. and A.~Lozano, \emph{Foundations of MIMO Communication}.\hskip 1em plus 0.5em minus 0.4em\relax Cambridge University Press, 2018.

\bibitem{matrix_transforms_optimization}
K.~Shen \emph{et~al.}, ``Optimization of {MIMO} device-to-device networks via matrix fractional programming: A minorization–maximization approach,'' \emph{IEEE/ACM Trans. Netw.}, vol.~27, no.~5, p. 2164–2177, 2019.

\bibitem{MSE_matrix}
S.~Gong \emph{et~al.}, ``Secure communications for dual-polarized {MIMO} systems,'' \emph{IEEE Trans. Signal Process.}, vol.~65, no.~16, pp. 4177--4192, 2017.

\bibitem{matrix_analysis}
X.-D. Zhang, \emph{Matrix Analysis and Applications}.\hskip 1em plus 0.5em minus 0.4em\relax Cambridge University Press, 2017.

\bibitem{Riemmanian_optimization_Kostas}
G.~C. Alexandropoulos \emph{et~al.}, ``Counteracting eavesdropper attacks through reconfigurable intelligent surfaces: A new threat model and secrecy rate optimization,'' \emph{IEEE Open J. Commun. Soc.}, vol.~4, pp. 1285--1302, 2023.

\bibitem{Riemannian_opt_steps}
J.~Wu \emph{et~al.}, ``Energy-efficient power control and beamforming for reconfigurable intelligent surface-aided uplink {IoT} networks,'' \emph{IEEE Trans. Wireless Commun.}, vol.~21, no.~12, pp. 10\,162--10\,176, 2022.

\bibitem{George_interference}
G.~C. Alexandropoulos and C.~B. Papadias, ``A reconfigurable iterative algorithm for the {$K$}-user {MIMO} interference channel,'' \emph{Elsevier Signal Process.}, vol.~93, no.~12, pp. 3353--3362, 2013.

\bibitem{Zhang_MIMO_IRS}
S.~Zhang and R.~Zhang, ``Capacity characterization for intelligent reflecting surface aided {MIMO} communication,'' \emph{IEEE J. Sel. Areas Commun.}, vol.~38, no.~8, pp. 1823--1838, 2020.

\bibitem{GR_001}
\BIBentryALTinterwordspacing
{ETSI GR RIS 001}, ``{Reconfigurable Intelligent Surfaces (RIS); Use cases, deployment scenarios and requirements},'' Apr. 2023. [Online]. Available: \url{https://portal.etsi.org/webapp/WorkProgram/Report_WorkItem.asp?WKI_ID=63938}
\BIBentrySTDinterwordspacing

\bibitem{ETSI_THz_2023}
\BIBentryALTinterwordspacing
{ETSI GR THz 001}, ``Tera{H}ertz modeling ({THz}); identification of use cases for {THz} communication systems,'' 2024. [Online]. Available: \url{https://www.etsi.org/deliver/etsi_gr/THz/001_099/001/01.01.01_60/gr_THz001v010101p.pdf}
\BIBentrySTDinterwordspacing

\bibitem{AJG2024}
G.~C. Alexandropoulos \emph{et~al.}, ``Characterization of indoor {RIS}-assisted channels at $304$ {GHz}: Experimental measurements, challenges, and future directions,'' \emph{arXiv preprint:2412.07359}, 2024.

\end{thebibliography}

\end{document}